\documentclass[12pt,a4paper,english]{article}
\usepackage{titling}
\usepackage[affil-it]{authblk}
\usepackage{longtable}
\usepackage{lscape}
\usepackage{graphicx}
\usepackage{epsfig}
\usepackage{dsfont}
\usepackage{amsfonts}
\usepackage{amsmath}
\usepackage{amsthm}
\usepackage{amssymb}
\usepackage{bbold}
\usepackage[english]{babel}
\usepackage[utf8]{inputenc}
\usepackage[left=0.8in,right=0.8in,top=1.0in,bottom=1.2in]{geometry}
\usepackage{calc}
\usepackage{cancel}
\usepackage{float}
\usepackage{slashed}
\usepackage{mathrsfs}
\usepackage{pdfpages}
\usepackage{tabu}
\usepackage{simplewick}
\usepackage[hidelinks]{hyperref}
\usepackage{bm}
\usepackage{multirow}
\usepackage{hhline}
\usepackage{pifont}
\usepackage{float}
\usepackage{xcolor}
\usepackage[normalem]{ulem}
\usepackage{cite}
\usepackage{caption}
\usepackage{subcaption}

\usepackage{booktabs}

\numberwithin{equation}{section}
\numberwithin{figure}{section}
\numberwithin{table}{section}

\makeatletter
\g@addto@macro\bfseries{\boldmath}
\makeatother

\allowdisplaybreaks
\usepackage{ulem}
\makeatletter
\def\fmslash{\@ifnextchar[{\fmsl@sh}{\fmsl@sh[0mu]}}
\def\fmsl@sh[#1]#2{%
  \mathchoice
    {\@fmsl@sh\displaystyle{#1}{#2}}%
    {\@fmsl@sh\textstyle{#1}{#2}}%
    {\@fmsl@sh\scriptstyle{#1}{#2}}%
    {\@fmsl@sh\scriptscriptstyle{#1}{#2}}}
\def\@fmsl@sh#1#2#3{\m@th\ooalign{$\hfil#1\mkern#2/\hfil$\crcr$#1#3$}}
\makeatother
\usepackage{graphicx}
\usepackage{xcolor}

\newcommand{\mBc}{{m_{B_c}}}
\newcommand{\mBcsq}{{m_{B_c}^2}}

\newcommand{\mpsi}{{m_{J/\psi}}}
\newcommand{\mpsisq}{{m_{J/\psi}^2}}

\def\eq#1{{Eq.~(\ref{#1})}}
\def\eqs#1#2{{Eqs.~(\ref{#1})--(\ref{#2})}}
\def\fig#1{{Fig.~\ref{#1}}}
\def\figs#1#2{{Figs.~\ref{#1}--\ref{#2}}}
\def\Table#1{{Table~\ref{#1}}}

\def\app#1{{Appendix~\ref{#1}}}

\newcommand{\DS}[1]{/\!\!\!#1}

\newcommand{\ccdot}{\!\cdot\!}

\begin{document}
\begin{titlepage}
\begin{flushright}
SI-HEP-2022-26\\
P3H-22-094\\
CERN-TH-2022-144 \\
\end{flushright}

\vspace{1.2cm}
\begin{center}
{\Large\bf
New Sum Rules for the \boldmath $B_c \to J/\psi $ \unboldmath Form Factors}
\end{center}
\begin{center}
{\sc  M.~Bordone,$^a$ A.~Khodjamirian,$^b$ Th.~Mannel$^b$} \\[3mm]
{\it $^a$ Theoretical Physics Department, CERN, 1211 Geneva 23, Switzerland}\\[0.3cm]
{\it $^b$
Center for Particle Physics Siegen (CPPS), Theoretische Physik 1\,,\\
Universit\"at Siegen, D-57068 Siegen, Germany 
}
\end{center}

\vspace{0.5cm}
\begin{abstract}
\vspace{0.2cm}\noindent
We derive new sum rules for the  form factors of the $B_c\to J/\psi\, \ell \bar{\nu}_\ell$ semileptonic transitions,
employing the vacuum-to-$B_c$ correlation function of the  $J/\psi$-interpolating and $b\to c$ weak currents.
In the heavy quark limit and at a space-like momentum transfer to the weak current,
a local operator-product expansion is valid
for this correlation function. As a result, in the leading power, the 
non-perturbative input is reduced to the decay constant of $B_c$ meson.
Furthermore, applying hadronic dispersion relation in the $J/\psi$ channel, we find that a non-vanishing OPE spectral density in the duality interval of $J/\psi$ emerges only at $\mathcal{O}(\alpha_s)$. We calculate this density in the relevant kinematic regime. The $B_c\to J/\psi$ form factors at space-like momentum transfer are then calculated from the new sum rules.


\end{abstract}

\end{titlepage}

\newpage
\pagenumbering{arabic}

\section{Introduction}

The semileptonic $B_c\to J/\psi\, \ell \bar\nu_\ell$ decay
was the discovery channel \cite{CDF:1998ihx} of the $B_c$ meson and 
still attracts a lot of interest. The main reason is
that the measured ratio \cite{LHCb:2017vlu} of the branching fractions with $\ell=\tau$ and $\ell=\mu$ reveals some tension with Standard Model (SM). Furthermore, the branching ratio of $B_c\to J/\psi\, \mu \bar\nu$ decay is a key input in the measurement of the $B_c$ fragmentation fraction \cite{LHCb:2019tea}. To obtain the exclusive semileptonic widths and their ratio,  one needs the $B_c\to J/\psi$ hadronic form factors
in the whole region of momentum transfer squared, $0\leq q^2\leq (m_{B_c}-m_{J/\psi})^2$. Recently, these form factors have been calculated using lattice QCD \cite{Harrison:2020gvo,Harrison:2020nrv} with an appreciable accuracy, thereby challenging all previous model calculations based on continuum QCD.
A specific property of the $B_c\to$~charmonium transitions is 
the fact that all valence quarks are heavy. This, generally,  enables 
us to represent   the 
$B_c\to J/\psi$ form factors in terms
of an overlap of heavy quarkonia wave functions. Along these lines, a broad variety of approaches 
was applied in the past,
from a  non-relativistic quark model\footnote{An instructive analysis of $B_c\to$ charmonium transitions, combining non-relativistic model with QCD factorization can be found in \cite{Bell:2005gw}} to non-relativistic QCD
(see e.g., \cite{Qiao:2011yz,Zhu:2017lqu}).
 Here we concentrate on the QCD sum rule approach, where 
the hadronic form factors are accessed indirectly, by matching a certain correlation function of quark currents to its hadronic dispersion relation.

In the past, three-point QCD sum rules were  applied to calculate the $B_c\to J/\psi$ form factors
(see e.g.,\cite{Colangelo:1992cx,Kiselev:1993ea,Kiselev:1999sc}). In this method, the correlation function represents a vacuum average  of a product 
of the $B_c$\,- and $J/\psi$\,-interpolating currents with the weak $b\to c$ current. 
This product of currents is expanded in local operators, including the perturbative and gluon-condensate contributions. 
The result is matched to the double dispersion relation in the external momenta of the interpolating currents.
The $B_c\to J/\psi$ transition enters the ground-state as a double pole in this dispersion relation, and the form factors are obtained by applying quark-hadron duality.  However, all previous analyses of the three-point sum rules were limited, in their perturbative part, to a computation of the leading-order (LO) contribution, which is a simple triangle quark loop without gluons. A calculation of the next-to-leading order (NLO) gluon radiative corrections to this triangle loop is a very demanding and probably not even doable task.  We note that using only the LO contribution is somewhat counter-intuitive since, at least for small $q^2$, a large momentum has to be transferred to the heavy spectator quark, which requires to exchange of one or more gluons.

Thus, it is desirable to look for an alternative sum rule. Let us note that the
standard technique of light-cone sum rules (LCSRs) used for the $B\to$ light-meson form factors can hardly be extended to the $B_c\to J/\psi$ transition. The reason is that keeping $m_c$ finite, one needs to define a light-cone distribution amplitude (DA) for a $J/\psi$ state. Such a definition cannot avoid  ill-defined $O(m_c^2x^2)$ terms  in the expansion near the light-cone $x^2\sim 0$.

In this paper, we suggest a new type of QCD sum rule for the $B_c\to J/\psi$ form factors.
The starting object is a correlation function in which the $B_c$ 
state is on the mass shell and the 
vector charmonium state is interpolated by the $\bar{c}\gamma_\mu c$ current. 
Our key observation is that, at a proper choice of external momenta, 
it is possible to expand the operator product of this correlation function in local operators. Thus,
since the spectator quark in the  $B_c \to J/\psi$ transition is heavy, we encounter a simpler operator product expansion (OPE)
than for correlation functions with the $B_{u,d,s}$-meson on-shell state. We remind that in the latter case, one necessary deals with a light-cone OPE\cite{Khodjamirian:2005ea,Khodjamirian:2006st,Faller:2008tr}, where the non-perturbative input consists of a set of 
the $B$-meson light-cone distribution amplitudes (DAs), defined in the Heavy Quark Effective Theory (HQET). In the case of the $B_c$ meson,  
the non-perturbative input in the heavy quark limit is reduced to a single hadronic parameter - the $B_c$ decay constant. 

Another important feature of the sum rules obtained in this paper concerns a nontrivial implementation of the quark-hadron duality. Applying a dispersion relation in the external momentum squared $p^2$ in the vector charmonium channel, one usually attributes to the ground-state  $J/\psi$-meson an interval of the OPE spectral density adjacent to the $\bar{c}c$ threshold,
that is, $p^2 \gtrsim 4m_c^2$.
For the correlation function of our choice and considering the region $q^2\lesssim 0 $, where the OPE is valid, the spectral density vanishes at $\mathcal{O}(\alpha_s^0)$ near $p^2\sim \,4m_c^2$ and is nonzero only at much larger values of $p^2$.
The spectral density saturating the $J/\psi$ duality interval starts 
only at $\mathcal{O}(\alpha_s)$, reflecting the fact that a hard gluon is 
needed for a momentum transfer to the spectator quark, at least, at small and negative values of $q^2$.
We find that the spectral density in the duality interval 
is reduced to the two specific cut diagrams, which are computed applying a standard Cutkosky rule.

The paper is organized as follows. In Section~\ref{sect:correlSR} we introduce the correlation function
and derive the sum rules for the $B_c\to J/\psi$ form factors in  two different versions: the Borel-transformed and power-moment ones. 
In Section~\ref{sect:locOPE} we demonstrate the validity of the local OPE and obtain the LO
expression for the correlation function.
Section~\ref{sect:impart} is devoted to the analytical properties of the OPE diagrams. 
In Section~\ref{sect:NLO} we compute the $\mathcal{O}(\alpha_s)$ spectral density.
The numerical analysis is presented in Section ~\ref{sect:Numz} and in 
Section~\ref{sect:disc} we conclude. 
The appendices contain the  
  calculation of the master integrals for the cut diagrams (Appendix \ref{app:Im}),  
 the formulae  for tensor integrals (Appendix \ref{app:Tens} ) and  
some bulky expressions for the diagrams with axial current (Appendix  \ref{app:axi}).

\section{Correlation function and sum rule}
\label{sect:correlSR}
We start from the correlation function defined as
\begin{equation}
\begin{aligned}
F_{\mu\nu}(p,q)&=i\int d^4x\,e^{ipx}\langle 0|T\{\bar{c}(x)\gamma_\mu 
c(x)\bar{c}(0)\Gamma_\nu b(0)\} |\bar{B}_c(p+q)\rangle
\\
&=\epsilon_{\mu\nu \alpha\beta} 
q^\alpha p^\beta F^V(p^2,q^2)+ F^A_{\mu\nu}(p,q)\,,
\end{aligned}
\label{eq:corr}
\end{equation}
where the current $\bar{c}\gamma_\mu c$ interpolates the $J/\psi$ and other 
$\bar{c}c$ hadronic states with $J^P=1^-$ . This current forms a time-ordered product with
the $b\to c$ weak current, 
where $\Gamma_\nu=\gamma_\nu(1-\gamma_5)$.
The product of currents is inserted between the on-shell $\bar{B}_c$\,-\,meson state and the vacuum.
The correlation function $F_{\mu\nu}$ in \eq{eq:corr} is then separated into two terms: the one proportional to $F^V$  corresponds to an insertion of the weak vector current, while $F^A_{\mu\nu}$ corresponds to the weak axial-vector current. The latter term is decomposed into Lorentz-invariant quantities as 
\begin{equation}
\begin{aligned}
F_{\mu\nu}^A(p,q) =\,& g_{\mu\nu}F_{(g)}^A(p^2,q^2)
+q_{\mu}q_{\nu}F_{(qq)}^A(p^2,q^2)
+q_\mu p_\nu F_{(qp)}^A(p^2,q^2)\\
+\,&p_\mu q_\nu F_{(pq)}^A(p^2,q^2)
+p_\mu p_\nu F_{(pp)}^A(p^2,q^2)
\,.
\end{aligned}
\label{eq:FAdecomp}
\end{equation}
Since the current $\bar{c}\gamma_\mu c$ is conserved, the correlation function $F_{\mu\nu}$ vanishes after multiplying it by $p^\mu$
(neglecting possible contact terms).
For the vector-current part in Eq.(\ref{eq:corr})
this property is fulfilled automatically. For the axial-current part the equation $p^\mu F^A_{\mu\nu}=0$ results in two additional relations between the five invariant amplitudes entering (\ref{eq:FAdecomp}). These relations are obtained by putting to zero the coefficients at $p_\nu$ and $q_\nu$ in $p^\mu F^A_{\mu\nu}$. They, however, have no impact  on the sum rules we obtain below from 
the axial-current part. The reason is that all the three
invariant amplitudes that we use remain
independent after imposing the current conservation
condition.

In what follows, we use standard definitions of
the $B_c\to J/\psi$ form factors for the vector and axial-vector currents\,\footnote{Throughout this paper we use the convention $\epsilon_{0123}=-\epsilon^{0123}=1$.}: 
\begin{equation}
\langle J/\psi (p,\varepsilon) | \bar{c} \gamma_\nu b |
\bar{B}_c(p+q) \rangle = \epsilon_{\nu\rho \alpha\beta} 
\varepsilon^{*\rho} q^\alpha p^\beta \frac{2V(q^2)}{m_{B_c}+m_{J/\psi} }, 
\label{eq:Vdef}
\end{equation}
\begin{equation}
\begin{aligned}
&\langle J/\psi (p,\varepsilon) | \bar{c} \gamma_\nu\gamma_5 b |
\bar{B}_c(p+q) \rangle = i\Bigg(\varepsilon_\nu^*-\frac{(\varepsilon^*\cdot q) q_\nu}{q^2}\Bigg)(m_{B_c}+m_{J/\psi})A_1(q^2)
\\
&-i(\varepsilon^*\cdot q)\Bigg((2p+q)_\nu-\frac{m_{B_c}^2-m_{J/\psi}^2}{q^2}q_\nu\Bigg)\frac{A_2(q^2)}{m_{B_c}+m_{J/\psi}} 
+i (\varepsilon^*\cdot q)q_\nu\frac{2m_{J/\psi}}{q^2}A_0(q^2),
\end{aligned}
\label{eq:Aff}
\end{equation}
where the following endpoint relation applies
\begin{equation}
A_0(0)=\frac{m_{B_c}+m_{J/\psi}}{2m_{J/\psi}}A_1(0)- 
\frac{m_{B_c}-m_{J/\psi}}{2m_{J/\psi}}A_2(0)\,.
\label{eq:A3A0}
\end{equation}
In addition, we introduce the helicity form factor
 \begin{equation}
     A_{12} = \frac{(\mBc+\mpsi)^2(\mBcsq-\mpsisq-q^2)A_1-\lambda(\mBcsq,\mpsisq,q^2) A_2}{16 \mBc \mpsisq(\mBc+\mpsi)}\,,
\label{eq:a12}
\end{equation}
 where $\lambda(a,b,c)= a^2+b^2+c^2-2ab -2ac-2bc$  is the  K\"allen function.

The sum rules that we derive here are based on the hadronic dispersion relation 
for the correlation function defined in \eq{eq:corr},
considered as an analytic function in the variable $p^2$ at fixed $q^2$. Inserting in \eq{eq:corr} the total set of the $\bar{c}c$ hadronic states with $J^P=1^-$, we isolate the ground  $J/\psi$-state contribution to the dispersion relation:
\begin{eqnarray}
 &&F_{\mu\nu}^{(J/\psi)}(p,q)=\frac{\langle 0 | \bar{c}\gamma_\mu c\overline{| J/\psi (p,\varepsilon)\rangle \langle J/\psi (p,\varepsilon)} | \big(\bar{c} \gamma_\nu b - \bar{c} \gamma_\nu\gamma_5 b \big)|\bar{B}_c(p+q) \rangle}{m_{J/\psi}^2-p^2}
 \nonumber\\
 &&=\frac{m_{J/\psi}f_{J/\psi}}{m_{J/\psi}^2-p^2}
\Bigg\{
\epsilon_{\mu\nu \alpha\beta} 
q^\alpha p^\beta \frac{2V(q^2)}{m_{B_c}+m_{J/\psi}}
\nonumber\\
&& +
\left(g_{\mu\nu}-\frac{q_\mu q_\nu}{q^2}+ \frac{(p\cdot q)p_\mu q_\nu}{q^2}-
\frac{p_\mu p_\nu}{m_{J/\psi}^2}\right)(m_{B_c}+m_{J/\psi})iA_1(q^2)
\nonumber\\
&&-\left(q_\mu-\frac{(p\cdot q) p_\mu}{m_{J/\psi}^2}\right)
\left((2p+q)_\nu -\frac{m_{B_c}^2-m_{J/\psi}^2}{q^2}q_\nu\right)
\frac{ iA_2(q^2)}{m_{B_c}\!+\!m_{J/\psi}} 
\nonumber\\
&&+\left(q_\mu q_\nu -\frac{(p\cdot q)p_\mu q_\nu}{m_{J/\psi}^2} \right)
\frac{2m_{J/\psi}}{q^2}iA_0(q^2)\Bigg\}.
\label{eq:Affv}
\end{eqnarray}
where the decay constant of the $J/\psi$ is defined as
$\langle 0|\bar{c}\gamma_\mu c |J/\psi(p,\varepsilon)\rangle= \varepsilon_\mu m_{J/\psi}f_{J/\psi} $, and the overline indicates summation over the $J/\psi$ polarizations.

Comparing the above expression with the Lorentz-decomposition of 
the correlation function (\ref{eq:corr}), it is possible to  
obtain a hadronic dispersion relation for each separate Lorentz-invariant amplitude.
For the vector-current part, we have:
\begin{eqnarray}
F^V(p^2,q^2)=\frac{2m_{J/\psi}f_{J/\psi}V(q^2)}{(m_{B_c}+m_{J/\psi})(m_{J/\psi}^2-p^2) }+
\int\limits_{s_h}^\infty \!\! ds \,\frac{\rho_h^V(s,q^2)}{s-p^2}\,,
\label{eq:Vdisp}
\end{eqnarray}
where the integral over $\rho_h^V$ includes the contributions
of hadronic states with $c\bar{c}$-content\footnote{Strictly speaking, the hadronic spectral density in the $\bar{c}\gamma_\mu c$ 
channel includes also light quark-antiquark  states, but their contributions are strongly 
suppressed as manifested by  a very small  width of the $J/\psi$
annihilation to light hadrons. 
In terms of duality, these contributions to $\rho^V_h$ 
would correspond to diagrams with additional virtual gluon lines and quark loops, suppressed with multiple powers of $\alpha_s$.}, 
which are heavier than $J/\psi$  and have spin-parity $J^P=1^-$.
We ignore possible subtractions in \eq{eq:Vdisp} since we will perform the Borel transform or multiple differentiation.

In the following sections, we calculate the correlation function in terms of an OPE
and obtain the functional dependence of the invariant amplitudes on  $p^2$ and $q^2$.
The result of the OPE calculation 
will then be converted  into a dispersive form. For the amplitude $F^V$, 
the dispersion relation reads:
\begin{eqnarray}
F^{V(OPE)}(p^2,q^2)= \frac1{\pi}\int\limits_{s_{min}}^\infty\! ds \,\frac{\mbox{Im}F^{V(OPE)}(s,q^2)}{s-p^2}\,,
\label{eq:OPEdisp}
\end{eqnarray}
where $s_{min}$ is the 
lowest threshold of the quark-level diagrams contributing to the OPE.

The next step, indispensable for any QCD sum rule, is the use of quark-hadron duality. 
The integral over $\rho^V$ in the dispersion relation (\ref{eq:Vdisp})
is approximated by the part of the integral (\ref{eq:OPEdisp}) above an effective threshold $s_0$. Substituting \eq{eq:OPEdisp}
 in \eq{eq:Vdisp}, we subtract from both sides the integrals that are equal due to duality. 
Based on the experience with the  original two-point
QCD sum rules for charmonium \cite{Shifman:1978bx,Shifman:1978by}, we assume that the interval of the OPE spectral density dual to the $J/\psi$ contribution in \eq{eq:Vdisp}
spans from the $c\bar{c}$ threshold $s_{min}=4m_c^2$ to 
$s_0=(2m_c+\omega_0)^2$ where the energy interval $\omega_0$ does not scale with the heavy mass $m_c$. 
Implementing duality and applying the Borel transform in the variable $p^2$, we obtain
 the desired sum rule for the vector form factor:
 \begin{eqnarray}
\frac{2 m_{J/\psi}f_{J/\psi}V(q^2)}{(m_{B_c}+m_{J/\psi})}e^{-m_{J/\psi}^2/{\cal M}^2}=
 \frac{1}{\pi}\int\limits_{4m_c^2}^{s_0}\! ds \,e^{-s/{\cal M}^2}\mbox{Im}F^{V(OPE)}(s,q^2)\,.
\label{eq:VBorelSR}
\end{eqnarray}

A procedure similar to the one used to obtain the above sum rule is repeated for the axial-current form factors.
Comparing the coefficients associated with the same Lorentz structures in \eq{eq:FAdecomp} and \eq{eq:Affv}, we find that
the form factors $A_1$ and $A_2$ are the only ones multiplied by  the structures 
$g_{\mu\nu}$ and $q_\mu p_\nu$, respectively. Equating the invariant 
coefficient of each of these structures in the OPE expression to its counterpart in 
the hadronic dispersion relation, we obtain  the following two sum rules: 
 \begin{eqnarray}
&&m_{J/\psi}(m_{B_c}\!+\!m_{J/\psi})f_{J/\psi}iA_1(q^2)e^{-m_{J/\psi}^2/{\cal M}^2}\!=\!
 \frac{1}{\pi}\int\limits_{4m_c^2}^{s_0}\! ds \,e^{-s/{\cal M}^2}\mbox{Im}F^{A(OPE)}_{(g)}(s,q^2)\,,
\label{eq:A1ff}
\\
&&\frac{2m_{J/\psi}}{(m_{B_c}+m_{J/\psi})}f_{J/\psi}iA_2(q^2)e^{-m_{J/\psi}^2/{\cal M}^2}\!=\!
 -\frac{1}{\pi}\!\int\limits_{4m_c^2}^{s_0}\! \! ds \,
 e^{-s/{\cal M}^2}\,\mbox{Im}F_{(qp)}^{A(OPE)}(s,q^2)\,.
\label{eq:A2ff}
\end{eqnarray}
Finally, using the invariant amplitude multiplying the structure $q_\mu q_\nu$, 
we obtain an additional sum rule 
 \begin{eqnarray}
&&m_{J/\psi}f_{J/\psi}\Bigg[ 2m_{J/\psi} iA_0(q^2)
-(m_{B_c}+m_{J/\psi})iA_1(q^2) 
\nonumber\\
&&+\left(m_{B_c}^2\!-\!m_{J/\psi}^2-q^2\right)
\frac{iA_2(q^2)}{m_{B_c}\!+\!m_{J/\psi}}\Bigg]
e^{-m_{J/\psi}^2/{\cal M}^2}
\!\!=\!\frac{q^2}{\pi}\!\int\limits_{4m_c^2}^{s_0}\! ds 
\,e^{-s/{\cal M}^2}\mbox{Im}F^{A(OPE)}_{(qq)}(s,q^2)
\label{eq:A012ff}
\end{eqnarray}
which yields the form factor $A_0$, provided  $A_{1}$ and $A_2$ are determined from 
the sum rules in \eqs{eq:A1ff}{eq:A2ff}.
In addition, the helicity form factor $A_{12}$ defined in \eq{eq:a12}
is obtained as a linear
combination of the same sum rules. 

As an alternative method, we  consider the power-moment version of the new QCD sum rules. This version was more frequently used for charmonium, starting from the original two-point sum rules \cite{Shifman:1978bx,Shifman:1978by}.  
Taking as an example the vector form factor, we obtain the $n$-th power moment, 
differentiating both parts of the duality-subtracted dispersion relation (\ref{eq:Vdisp}) 
 $n$ times over $p^2$ at fixed $p^2\equiv -P^2\leq 0$ 
\footnote{ We remind that both versions of sum rules are related via limiting
transition: the Borel transform is obtained taking in the power moment 
the simultaneous limit $n\to \infty$ and $p^2\to-\infty$ and keeping the ratio $(-p^2/n)$ 
finite and equal to ${\cal M}^2$.}:
 \begin{eqnarray}
\frac{2m_{J/\psi}f_{J/\psi}V(q^2)}{(m_{B_c}+m_{J/\psi})(m_{J/\psi}^2+P^2)^{n+1}}=
 \frac{1}{\pi}\int\limits_{4m_c^2}^{s_0}\! \frac{ds}{(s+P^2)^{n+1}} \mbox{Im}F^{V(OPE)}(s,q^2)\,.
\label{eq:mom}
\end{eqnarray}
It is then straightforward to write down similar power-moment versions for the sum rules
(\ref{eq:A1ff})-(\ref{eq:A012ff}), replacing in the latter 
the exponential factor  $e^{-m_{J/\psi}^2/{\cal M}^2}$ ($e^{-s/{\cal M}^2})$ on l.h.s. (r.h.s.)  
by a factor $1/(m_{J/\psi}^2+P^2)^{n+1}$ ($1/(s+P^2)^{n+1}$).

\section{Validity of the local OPE}
\label{sect:locOPE}
At first sight, the correlation function (\ref{eq:corr}) does not essentially
differ from the ones introduced in Refs.~\cite{Khodjamirian:2005ea,Khodjamirian:2006st,Faller:2008tr} 
to obtain sum rules for the $B$-meson semileptonic transitions to light
and charmed mesons, respectively.
In these sum rules, the light-cone OPE was employed and the correlation functions
were calculated in terms of  universal $B$-meson distribution amplitudes (DAs)
defined in HQET. In all other aspects, the method largely followed the original  LCSRs developed in\cite{Balitsky:1986st,Balitsky:1989ry,Chernyak:1990ag}. 

Here, our main observation is that a  heavy  spectator $c$-quark in the initial $B_c$ state 
cardinally changes the situation for the correlation function (\ref{eq:corr}) and 
validates a simpler OPE in terms of local operators.
Consequently, there is no need to introduce  
distribution amplitudes describing the $B_c$ meson. As we shall see, the non-perturbative input in the sum rule (\ref{eq:VBorelSR}) at the leading 
power consists of a single parameter - the decay constant of the $B_c$ meson. 
The latter is calculated in lattice QCD or using conventional two-point QCD sum rules. The decay constant can, in principle, also be measured in the leptonic decays of the $B_c$, provided the CKM parameter $V_{cb}$ is known independently.

In what follows, we systematically employ the heavy quark limit for the correlation function in \eq{eq:corr}, assuming
\begin{equation}
m_b, \,  m_c\gg \bar{\Lambda}\sim \Lambda_{QCD}\,,
\label{eq:scales}
\end{equation}
where $\bar{\Lambda}$ is a typical binding energy in
a heavy quarkonium state such as  the $B_c$ or the $J/\psi$. 
Our goal is to 
formulate the method and perform calculations in the leading power approximation. Hence, we neglect all 
$\mathcal{O}(\bar{\Lambda}/m_c,\bar{\Lambda}/m_b)$ effects, 
so that
\begin{equation}
m_{B_c}\simeq m_b+m_c\equiv M, ~~~p_{B_c}=p+q\simeq m_bv+m_cv=Mv\,,
\label{eq:masses}
\end{equation}
where $v$ is the four-velocity of the $B_c$, with
$v=(1,\vec{0}\,)$ in the rest frame. 
Simultaneously, the virtualities   
of the interpolating and weak currents, respectively, $p^2$ and $q^2$,  
are chosen far from any hadronic threshold, assuming
\begin{equation}
p^2\ll 4m_c^2\,,~~~~~q^2\ll M^2. 
\label{eq:ineq}
\end{equation}

Under these conditions, a virtual $c$-quark
emitted and absorbed between the point $x$ and the origin in the correlation function (\ref{eq:corr})
is far off shell. 
\begin{figure}[t]
\centerline{
\includegraphics[scale=1.2]{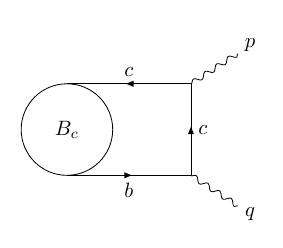}}
	\caption{The leading order diagram representing the correlation
 function (\ref{eq:corr}).}
	\label{fig:diagLO}  
\end{figure} 
Hence, at LO the correlation function is described 
by the diagram in Fig.~\ref{fig:diagLO}, with  a free $c$-quark propagator. 
The resulting expression is:
\begin{equation}
F^{(LO)}_{\mu\nu}(p,q)
=i^2\!\int\!\frac{d^4r}{(2\pi)^4}\big[\gamma_\mu\frac{\DS r+m_c}{r^2-m_c^2}\Gamma_\nu\big]_{\alpha\beta}\!\int \!d^4x\,e^{i(px-rx)}\langle 0|\bar{c}_\alpha(x) b_\beta(0) |\bar{B}_c(p+q)\rangle\,.
\label{eq:corr2}
\end{equation}

In the above, the hadronic matrix element of the non-local operator is factorized. 
To simplify the technical part of our
discussion, in the rest of this section we consider the vector part 
of the weak current, replacing $\Gamma_\nu$ by $\gamma_\nu$. 

The matrix element in (\ref{eq:corr2}) still depends on the masses of the
heavy quarks. In order to separate this dependence, we first perform a
redefinition of the heavy quark fields according to 
\begin{equation}
    c(x) = e^{i m_c (vx)} c_v(x) \quad \mbox{and} \quad b(x) = e^{-i m_b (vx)} b_v(x) \,,
\nonumber
\end{equation}
where $v=(1,\vec{0})$ defines the $B_c$ rest frame. Note that this corresponds to 
redefining the momenta of the charm-antiquark 
and bottom-quark as $p_c = m_c v+k_c$ and  $p_b = m_b v+k_b$, respectively.    

The redefined fields in the matrix element can now be expanded in $1/m_c$ and $1/m_b$  such that, respectively, 
\begin{equation}
    c_v(x) = h_v^\prime (x) + \cdots \qquad  \mbox{and} \quad b_v (x) = h_v (x) + 
    \cdots \,,
\nonumber
\end{equation}
where the ellipses denote terms of higher order in the inverse heavy mass. 

Thus the relevant matrix element can be written as  

\begin{equation}
\langle 0|\bar{c}_\alpha(x) b_\beta(0) |\bar{B}_c(p+q)\rangle 
=\exp(-im_cv\cdot x)\sqrt{m_{B_c}}\langle 0 | \bar{h}'_{v\alpha}(x)
h_{v\beta}(0) | \bar{B}_c(v)\rangle +O(\bar{\Lambda}/m_c,\bar{\Lambda}/m_b)\,,
\label{eq:repl}
\end{equation}
where the factor $\sqrt{m_{B_c}}\simeq\sqrt{M}$ reflects the normalization of the 
effective $B_c$ state.

Since 
we are dealing with quarkonium-like state, the dynamics 
encoded in the matrix element with the $h_v$ and $h_v^\prime$ fields can be described 
by the Schr\"odinger equation, which eventually 
takes care of the binding 
via the exchange of Coulomb gluons
\footnote{These effects can be calculated in the framework of NRQCD, 
introducing the small parameter of the heavy-quark relative velocity defined as  $v_c=|\vec{k_c}|(m_b+m_c)/(m_b m_c)$ in the rest frame of $B_c$ where $\vec{k}_c=-\vec{k_b}$.}.
However, we will only 
 work to leading order 
here, so that we do not need to dwell on these issues. The only point which is relevant for 
our analysis is that we may compute the hard 
contributions by matching the perturbative parts of the diagrams 
to the kinematics  with $p_b = m_b v$ and $p_c = m_c v$.

To proceed, we substitute
the relation (\ref{eq:repl}) in \eq{eq:corr2}, and further use
$$\gamma_\mu \DS r \gamma_\nu =-i\epsilon_{\mu\nu\rho\lambda}\gamma^\lambda\gamma_5r^\rho +...\, ,$$
where the ellipsis do not yield an $\epsilon$-tensor structure. This is also true for the terms in \eq{eq:corr2} that are proportional to $m_c$.
We obtain:
\begin{equation}
F^{(LO)}_{\mu\nu}(p,q)
=i\epsilon_{\mu\rho\nu\lambda}\sqrt{m_{B_c}}
\!\int\!\frac{d^4r}{(2\pi)^4}\frac{r^\rho}{r^2-m_c^2}\!
\int \!d^4x\,e^{i(p-mcv-r)x}
\langle 0 | \bar{h}'_{v}(x)\gamma^\lambda\gamma_5
h_{v}(0) | \bar{B}_c(v)\rangle\,.
\label{eq:corr2next}
\end{equation}

The next step is to expand the product of operators in 
the above hadronic matrix element in the series of 
local operators near $x=0$\,:
\begin{eqnarray}
\langle 0|\bar{h}'_v(x)\gamma^\lambda\gamma_5  h_{v}(0) |\bar{B}_c(v)\rangle= 
\sum\limits_{k=0}^\infty \frac{1}{k!}x^{\mu_1}x^{\mu_2}...x^{\mu_k}
\langle 0|\bar{h}'_v(0) \overleftarrow{D}_{\mu_1}\overleftarrow{D}_{\mu_2}...\overleftarrow{D}_{\mu_k}
\gamma^\lambda\gamma_5 h_{v}(0) |\bar{B}_c(v)\rangle\,,
\label{eq:exp3}
\end{eqnarray}
where for each term with $k$ derivatives, the following generic decomposition 
is valid:
\begin{equation}
\langle 0|\bar{h}'_v(0) \overleftarrow{D}_{\mu_1}\overleftarrow{D}_{\mu_2}...\overleftarrow{D}_{\mu_k}
\gamma^\lambda\gamma_5 h_{v}(0) |\bar{B}_c(v)\rangle=
i v^\lambda v_{\mu_1}v_{\mu_2}....v_{\mu_k}\Lambda^{(k)}_{B_c} + \dots\,.
\label{eq:xexpx}
\end{equation}
In the above, the ellipses denote structures 
containing $g_{\mu_i\mu_j}$ $(i,j=1,...k)$, that yield terms proportional to powers of  $x^2$ in the expansion (\ref{eq:exp3}).
Such terms generate contributions to \eq{eq:corr2}
that are suppressed at least by two powers of an inverse heavy-mass scale,
hence, we neglect them. We note in passing that
these contributions are of higher twist in the context of a more general light-cone expansion
near $x^2=0$ (see e.g., \cite{Colangelo:2000dp} for a detailed explanation). 

Furthermore, 
since the matrix elements (\ref{eq:xexpx}) 
are free of the heavy-quark mass  scales, 
their dimensionful parameters $\Lambda^{(k)}_{B_c}$ are proportional to
growing powers of the soft scale $\bar{\Lambda}$:

\begin{equation}
\Lambda_{B_c}^{(k)} \sim \bar{\Lambda}^{k}\hat{f}_{B_c}.
\label{eq:lamk}
\end{equation}

Here we use that  the parameter $\Lambda_{B_c}^{(0)}$ corresponding to the operator with the lowest dimension
in \eq{eq:xexpx} coincides with the static decay constant of $B_c$ meson defined via
\begin{equation}
\langle 0|\bar{h}'_v\gamma^\lambda\gamma_5 h_{v} |\bar{B}_c(v)\rangle=
i v^\lambda\hat{f}_{B_c}\,.
\label{eq:gk}
\end{equation}
The latter is related  to the decay constant in full QCD: 
$
\langle 0|\bar{c} \gamma^\lambda\gamma_5 b |\bar{B}_c(p+q)\rangle=
i (p+q)^\lambda f_{B_c}\,,
$
so that at leading power and at $\mathcal{O}(\alpha_s^0)$
\begin{equation}
f_{B_c}=\hat{f}_{B_c}/\sqrt{m_{B_c}}\,.
\label{eq:fBstat}
\end{equation}
Substituting \eqs{eq:exp3}{eq:lamk} in \eq{eq:corr2next}, we integrate over the coordinates, using 
\begin{eqnarray}
\int \!d^4x\,e^{i(p-mcv-r)x}
x^{\mu_1}x^{\mu_2}...x^{\mu_k}=
(-i)^k \frac{\partial}{\partial p_{\mu_1}}
\frac{\partial}{\partial p_{\mu_2}}
\frac{\partial}{\partial p_{\mu_3}}
...\frac{\partial}{\partial p_{\mu_k}}\big[\delta(p-m_cv-r)\big]\,.
\label{eq:intx}
\end{eqnarray}
A subsequent integration  over the four-momentum $r$ by means of the $\delta$-function yields:
\begin{equation}
F^{(LO)}_{\mu\nu}(p,q)
=-\frac{\epsilon_{\mu\nu\lambda\rho} v^\lambda p^\rho \sqrt{m_{B_c}}}{(p-m_cv)^2-m_c^2}\hat{f}_{B_c}
\Bigg\{1
+ \sum\limits_{k=1}^{\infty}\frac{(i)^k}{k !}
\bar{\Lambda}^{k}\Bigg[\frac{2(p-m_cv)\cdot v}{m_c^2-(p-m_cv)^2}\Bigg]^k
\Bigg\}\,.
\label{eq:LOansw}
\end{equation}

To specify the kinematics, we  choose the $B_c$-meson  rest frame in which
\begin{equation}
    v=\, (1,\vec{0}\,), ~~
    p =\, (p_0,0,0, |\vec{p}\,|)\,.
\label{eq:frame}
\end{equation}
In this frame, the components of the four-vector $p$ can be 
expressed via $q^2$ and $p^2$:
\begin{equation}
    p_0 =\frac{M^2+p^2-q^2}{2M}\,, \qquad |\vec{p}\,| = \frac{\sqrt{\lambda(M^2,p^2,q^2)}}{2M}\,,
\label{eq:kinem}
\end{equation}
where, retaining the leading power for the $B_c$ mass, we replace $m_{B_c}$ by $M$. 
Note that at $q^2=0$ these relations  simplify to 
\begin{equation} 
p_0 = \frac{1}{2M} (M^2+p^2)\,  \qquad |\vec{p}\,| = \frac{1}{2M} (M^2-p^2)\,.
\end{equation} 

Using for $p_0$  the  expression  in \eq{eq:kinem}, we obtain for the
coefficient multiplying the soft scale $\bar{\Lambda}$ in  \eq{eq:LOansw}:
\begin{equation}
\frac{2(p-m_cv)\cdot v}{m_c^2-(p-m_cv)^2}=
\frac{1}{m_c}\Bigg(1+\frac{p^2/M^2-2m_c/M}{1-q^2/M^2}\Bigg)
\Bigg(1-\frac{p^2m_b/(m_c M^2)}{1-q^2/M^2}\Bigg)^{-1}\,.
\label{eq:ratio}
\end{equation}
We make an important observation: for the scales and external momenta
 chosen according to the conditions  (\ref{eq:scales}) and (\ref{eq:ineq}),  
 the contributions with $k\geq 1 $  in 
\eq{eq:LOansw} are suppressed by powers of 
the parametrically small quantity $(\bar{\Lambda}/m_c)^k\ll 1$. The 
correlation function is dominated  by the lowest-dimensional local operator 
in the expansion (\ref{eq:exp3}). Neglecting the power-suppressed terms, we reduce \eq{eq:LOansw} to the hadronic matrix element of a single local operator,
determined by a known parameter - the $B_c$ decay constant.

Our final result
for the vector part of the correlation function (\ref{eq:corr}) at leading power in the inverse heavy-quark masses  and at LO in $\alpha_s$ directly reads off from \eq{eq:LOansw}:
\begin{equation}
 F^{V(LO)}(p^2,q^2)=\frac{f_{B_c}}{m_c^2-(p-m_cv)^2}=
 \frac{f_{B_c}M}{m_b\big(m_cM^2/m_b-q^2m_c/m_b-p^2\big)}\,,
\label{eq:LOfinal}    
\end{equation}
where we replace the static decay constant of $B_c$ by the full QCD decay
constant. For completeness, we present the axial-current part obtained 
from \eq{eq:corr2} in the same approximation, replacing $\Gamma_\nu $ by 
$-\gamma_\nu\gamma_5$:
\begin{equation}
 F^{A(LO)}_{\mu\nu}(p,q)=\frac{if_{B_c}M}{m_c^2-(p-m_c v)^2}\Big[(p\cdot v)g_{\mu\nu}-
 \big(v_\mu p_\nu+v_\nu p_\mu \big) +2m_c v_\mu v_\nu\Big]\,.
\label{eq:LOA}    
\end{equation}
Replacing $Mv=p+q$, we recover  the decomposition in \eq{eq:FAdecomp} in terms of the 
four-momenta  $p$ and 
$q$. 

Two additional comments are in order.
First, as explicitly follows from \eq{eq:ratio}, the suppression 
of higher-dimensional operators
becomes more effective at negative and large values of $q^2$. This will become
important for our numerical analysis below.

Second, the local OPE ceases to be valid if the valence $c$-quark in the correlation function is replaced by a light quark, hence substituting the on-shell $B_c$-state with the $B_{q}$-state, ($q=u,d,s$).
To see that, we perform an analogous expansion as in \eq{eq:LOansw}, assuming 
$m_q=0$ and find that  $p-m_cv$ has to be replaced by $p$. 
Consequently, the $\bar{\Lambda}/m_c$ suppression is effectively removed, being multiplied by a parametrically large $\mathcal{O}(m_b/m_c)$ factor. Hence, the whole tower  of local operators 
should be taken into account, and the hadronic matrix element  forms the $B_q$-meson 
light-cone DA at the leading twist. The light-cone expansion remains valid 
and we end up with the usual scheme of LCSRs   
for the $B_q\to D^*_q $ form factors (see, e.g. \cite{Faller:2008tr}). 
Suppose that  we replace also the virtual $c$-quark with a light quark,
correlating the $b\to q$ weak current with a light-quark current.
Then, instead of the  $\bar{\Lambda}/m_c$ , one has $\bar{\Lambda}m_b/(-p^2)$,
signaling again that the local expansion is not applicable. In this case,
the light-cone OPE is  valid at space-like and large $p^2$, yielding 
\cite{Khodjamirian:2005ea, Khodjamirian:2006st} LCSRs with $B_q$ meson DAs for the $B_q \to$ light-meson form factors.

\section{Analytical properties of OPE diagrams}
\label{sect:impart}
Here we discuss in more details the spectral density of the OPE diagrams 
contributing to the correlation function (\ref{eq:corr}).
We will identify  and select  the diagrams 
that have a  non-vanishing  imaginary part  in the variable $p^2=s$ within the 
interval
\begin{equation}
4m_c^2\leq s \leq s_0\sim (2m_c+\omega_0)^2\,,
\label{eq:dualint}
\end{equation}
dual to $J/\psi$.
Only these diagrams contribute to
the sum rules (\ref{eq:VBorelSR})-(\ref{eq:A012ff}). 

It is straightforward to analyse the spectral density of  the LO diagram in Fig.~\ref{fig:diagLO}.
To this end, we consider the vector-current part of the amplitude  written on r.h.s. of \eq{eq:LOfinal} in terms of invariant variables. The amplitudes 
entering the axial-current part in \eq{eq:LOA} have the same analytical properties. 
The LO amplitude has a simple pole in the variable $p^2$, located at 
\begin{equation}
p_{pole}^2=\frac{m_c}{m_b}\big(M^2-q^2)\,,
\label{eq:pole}
\end{equation}
hence its spectral density
is a delta-function:
\begin{equation}
\mbox{Im}F^{V(LO)}(s,q^2)\sim \delta(s-p_{pole}^2)\,.
\label{eq:impole}
\end{equation}
Note that the position (\ref{eq:pole}) of the pole  depends 
on $q^2$. Let us first consider  $q^2=0$, where we have
\begin{equation}
p^2_{pole} = m_cM^2/m_b\sim m_c m_b\,,
\label{eq:ppole}
\end{equation} 
which means that for $m_b \gg m_c$  
the LO spectral density (\ref{eq:impole}) vanishes within the interval (\ref{eq:dualint}).  
Thus, in the absence of energetic gluons, the final state in the $B_c\to c\bar{c} $ transition has an invariant mass,
which is parametrically much larger than $2m_c$, that is, far above the mass of the lowest charmonium resonance.
This is easy to understand in the rest frame of the initial $b\bar{c}$ state in which the $c$ quark originating from the weak decay 
of the $b$ quark at large recoil forms a large invariant mass with the spectator 
$\bar{c}$ quark. 
The distance between the pole in the LO spectral density and the characteristic duality interval
(\ref{eq:dualint})  increases  at $q^2<0$, where also the local 
OPE is expected to have a better convergence. Moreover, making  $-q^2$ sufficiently large, one can keep the pole far enough from the duality interval even at $m_b  \gtrsim m_c$.  
Note also that at the zero-recoil point,  $q^2=(m_b-m_c)^2$, the 
pole moves to $p_{pole}^2=4m_c^2$, and thus  inside 
the duality interval (\ref{eq:dualint}). However, this 
region of $q^2$ corresponds to the soft mechanism 
which is determined by the overlap of the $B_c$ and the $J/\psi$ wave functions, and
cannot be described by the OPE and sum rules we have set up. 

\begin{figure}
     \centering
     \begin{subfigure}[b]{0.45\textwidth}
         \centering
         \includegraphics[width=0.7\textwidth]{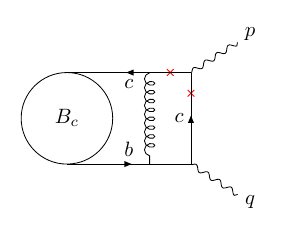}
         \caption{}
         \label{fig:diagNLOa}
     \end{subfigure}
     \hspace{5mm}
     \begin{subfigure}[b]{0.45\textwidth}
         \centering
         \includegraphics[width=0.7\textwidth]{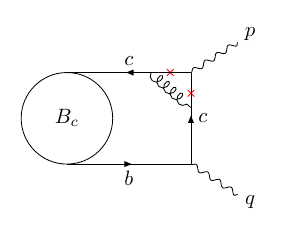}
         \caption{}
         \label{fig:diagNLOb}
     \end{subfigure}
     \newline
     \hfill
     \begin{subfigure}[b]{0.45\textwidth}
         \centering
         \includegraphics[width=0.7\textwidth]{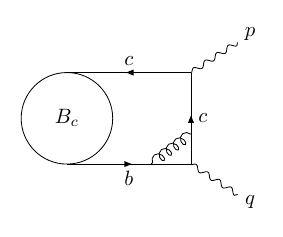}
         \caption{}
         \label{fig:diagNLOc}
     \end{subfigure}
     \hspace{5mm}
          \begin{subfigure}[b]{0.45\textwidth}
         \centering
         \includegraphics[width=0.7\textwidth]{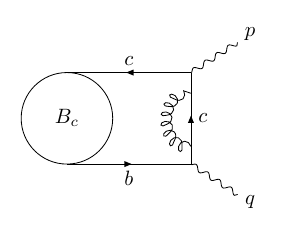}
         \caption{}
         \label{fig:diagNLOd}
     \end{subfigure}
        \caption{The NLO, $O(\alpha_s)$ diagrams: (a) the box, (b),(c)- the vertices and (d)-the self-energy of $c$-quark. Crosses indicate the cuts relevant
 for the sum rules.}
        \label{fig:diagNLO}
\end{figure}

The above discussion leads us to an important conclusion: at $q^2\leq 0$, the OPE spectral density 
dual to the $J/\psi$ contribution  
is non-vanishing only if a hard gluon is exchanged, and hence it arises 
at $O(\alpha_s)$. The corresponding diagrams are shown in Fig.~\ref{fig:diagNLO}.
A direct calculation of their imaginary part in $p^2$ is possible, 
 using the unitarity relation for the perturbative diagrams.
Applying a standard Cutkosky rule to each diagram, one can compute 
the contributions of all possible cuts of quark and gluon lines with respect to the variable $p^2$. The results should be then summed up over all diagrams. 
However, this task is quite tedious, because, apart from the simple two-particle cuts,
there are also three-particle ones with a complicated phase space.
Moreover, in the course of this calculation, one usually encounters
spurious infrared singularities which should cancel in the sum of all cuts
\footnote{An alternative possibility is to calculate the one-loop diagrams 
in terms of Feynman integrals and analytically continue the resulting expression
in $p^2$. This is technically also not simple, having in mind the presence of four different scales: the heavy quark masses and external momenta $q$ and $p$.}.
It is however possible to substantially reduce the amount of calculations if we 
confine ourselves to those contributions to the NLO spectral density that are
non-vanishing in the duality interval  in \eq{eq:dualint}. 

In order to select the relevant diagrams and their cuts, it is instructive 
to discuss  general analytical properties of the OPE diagrams. Neglecting the binding energies of the $b$ and $\bar{c}$ quarks
in the heavy quark limit, 
we interpret each OPE diagram as an amplitude of $2\to 2$ scattering where  the initial state consists of the on-shell $b$ and $\bar{c}$ quarks and the final state consists of two external ``particles" with squared masses $p^2$ and $q^2$. 
In general, any such $2\to 2$  amplitude depends on six independent invariant variables:
$m_b^2,m_c^2,q^2,p^2,s,t$  where $s$ and $t$ are the
usual Mandelstam variables related to the total energy and momentum transfer squared, respectively. In our case  $s=(m_b+m_c)^2$ is fixed at the initial state threshold, and,
correspondingly, $t=(p-m_cv)^2$ is also fixed. One can 
check this with the relations for the 
boundaries of Mandelstam variables \cite{Kibble:1960zz}, valid also for space-like values of $q^2$ or $p^2$. For the momentum transfer squared  we obtain:
\begin{equation}
t=\frac{m_b}{M}p^2+\frac{m_c}{M}q^2-m_cm_b.
\label{eq:t}
\end{equation}
This equation is valid for all diagrams describing the correlation function
in the static approximation for the $B_c$ meson. 

The LO diagram in Fig.~\ref{fig:diagLO}, viewed as a scattering amplitude, represents a $t$-channel exchange of a $c$-quark. Hence, its expression  
in \eq{eq:LOfinal} depends only on the variable $t$.
But since $t$ itself depends on $p^2$ via \eq{eq:t}, there is
an induced $p^2$ dependence of $F^{V(LO)}$, which is
explicit on the r.h.s. of \eq{eq:LOfinal}. This dependence
is, in fact, the only source of the imaginary part (i.e. of pole singularity) of the LO diagram in the variable $p^2$. 

The analytical properties of the NLO diagrams in 
\fig{fig:diagNLO} are less trivial. We encounter a 
combination of a direct and induced dependence on $p^2$. 
Consider, e.g., the box diagram  in \fig{fig:diagNLOa} with a gluon exchange between the $b$ and $\bar{c}$ quarks. The  direct contribution 
to the imaginary part in the variable $p^2$ corresponds to 
putting on-shell (cutting)  the $c$ and $\bar{c}$ lines adjacent to the $c\bar{c}$ vertex. This part of the OPE spectral density 
starts at the threshold $p^2=4m_c^2$ and, hence, is non-vanishing in the duality interval (\ref{eq:dualint}) that we are interested in.

Another contribution to the imaginary part in $p^2$ originates from the same diagram and corresponds to cutting the gluon and $c$-quark lines.
Similar to the LO diagram, this contribution depends on $p^2$ indirectly
via the variable $t$, which in this case is the squared sum of 
the gluon and $c$-quark momenta. Denoting these on-shell momenta and their sum by $k$, $f$, and $l=k+f$, respectively, so that $k^2=0$, and $f^2=m_c^2$, we have $t=l^2$.
Since we are only interested in kinematics, it is possible to 
contract in the box diagram the remaining virtual $c$-quark
and $b$-quark lines into points, so that the cut diagram is effectively reduced to a two-point diagram with an external timelike momentum $l$ 
flowing from the lower to the upper vertices.
 Furthermore, in  the chosen rest frame of the $B_c$ and using the momentum 
conservation at the lower vertex, $l=m_bv-q$,
we can express the energy component of $l$ via $t$ and $q^2$ as
$ l_0=(m_b^2+t-q^2)/(2m_b)$.
On the other hand, from the momentum conservation in the upper vertex, it follows that $p=l+m_cv$, hence
\begin{equation}
p^2=t+2m_cl_0+m_c^2=t\left(1+\frac{m_c}{m_b}\right)+m_cM-q^2\frac{m_c}{m_b}\,,
\label{eq:p2}
\end{equation}
which is another form of the general relation (\ref{eq:t}).
Since the on-shell gluon momentum increases the invariant mass
of the $c$-quark - gluon pair, we always have $t\geq m_c^2$. Therefore, 
according to \eq{eq:p2}, at $q^2\lesssim 0$ the $p^2$ region where the quark-gluon cut has a non-vanishing contribution lies far above the duality interval and does not contribute to the sum rules.

Analyzing the vertex diagram in \fig{fig:diagNLOb} in a similar way,
we identify the second contribution to the OPE spectral
density that is non-vanishing in the duality interval. It corresponds 
to the $c\bar{c}$ cut adjacent to the  upper vertex. 
The quark-gluon cuts of this diagram and of the remaining NLO diagrams in \figs{fig:diagNLOc}{fig:diagNLOd} result in the contributions located 
above this interval. Finally, we notice 
that the cuts of a single $c$-quark line in 
the two latter diagrams should also not be taken into account, since they produce a simple pole in $p^2$ at the same position 
as for the LO diagram.

Summarizing, we are left with two contributions to the OPE spectral density
relevant for the sum rules, and they are given by  the diagrams 
with cuts indicated by crosses in \figs{fig:diagNLOa}{fig:diagNLOb}.
It is interesting to note that these contributions are in one-to-one correspondence with a simple picture 
where the wave functions of  the $B_c$ and the $J/\psi$ states
are convoluted with a hard kernel, which, to leading order, is the one-gluon exchange 
 (cf. \cite{Bell:2005gw}).
The two corresponding diagrams are obtained from the cut diagrams in 
\figs{fig:diagNLOa}{fig:diagNLOb}
if the $\bar{c}c$ vertex is replaced by the $J/\psi$ state.
From this simple point of view it becomes also clear, that this 
new sum rule is restricted to negative and possibly also to small positive values of $q^2$, since at $q^2_{\rm max}$ the exchanged gluon becomes soft.

\section{The OPE spectral density}
\label{sect:NLO} 
Here we calculate separately the two contributions to 
the OPE spectral density, stemming from  the $\bar{c}c$ cuts of the NLO box and vertex diagrams. 

\subsection{The $\bar{c}c$ cut of the box diagram}
First, we  shall obtain the full expression of the diagram in Fig.~\ref{fig:diagNLO}(a) in terms of a Feynman integral. Inserting in (\ref{eq:corr}) a propagator 
\begin{equation}
\langle 0 | T\{A^a_\rho(z) A^b_\omega(y)|0\rangle =-i\delta^{ab}D_{\rho\omega}(z-y)\,=-i\delta^{ab}g_{\rho\omega}\int \frac{d^4k}{k^2}e^{ik(z-y)}
\nonumber
\end{equation}
(in the Feynman gauge) of the gluon line 
between $b$ and $c$ quarks, and contracting the  quark fields
into free propagators 
\begin{equation}
\langle 0 | T\{Q^i_\alpha(z)\bar{Q}^k_\beta(x)|0\rangle =i \delta^{ik}S_{Q\,\alpha\beta}(z-x)\,, ~~(Q=c,b)\,,
\nonumber
\end{equation}
we obtain:
\begin{eqnarray}
F_{\mu\nu}^{(box)}(p,q)=16i\pi\alpha_s\int d^4x\,e^{ipx}
\int d^4y \int d^4z
\langle 0|\bar{c}_\alpha(z)b_\beta(y)|\bar{B}_c(p+q)\rangle
\nonumber\\
\big[\gamma_\rho S_c(z-x)\gamma_\mu S_c(x)\Gamma_\nu S_b(-y)\gamma_\lambda\big]_{\alpha\beta}D^{\rho\lambda}(z-y)  \,,
\label{eq:corr2q}
\end{eqnarray}
where Dirac indices are shown explicitly and the color trace is taken.
The hadronic matrix element entering the above expression is transformed with a coordinate translation\,,
\begin{eqnarray}
\langle 0|\bar{c}_\alpha(z)b_\beta(y)|\bar{B}_c(p+q)\rangle=
\langle 0|\bar{c}_\alpha(0)b_\beta(y-z)|\bar{B}_c(p+q)\rangle e^{-i(p+q)z}\,.
\label{eq:hme1}
\end{eqnarray}
After that,  we expand this matrix element, similarly to \eq{eq:repl}, 
in a HQET form
\begin{eqnarray}
\langle 0|\bar{c}_\alpha(0)b_\beta(y-z)|\bar{B}_c(p+q)\rangle 
&\simeq &e^{-im_bv(y-z)}\sqrt{m_{B_c}}\langle 0|\bar{h}'_\alpha(0)h_\beta(0)|\bar{B}_c(v)\rangle
\nonumber\\
&\simeq &-\frac{i}{4}e^{-im_bv(y-z)}f_{B_c}m_{B_c}\big[(1+\fmslash{v})\gamma_5\big]_{\beta\alpha}\,,
\label{eq:hme2}
\end{eqnarray}
and  adopt the local limit. More specifically,  
we neglect the contributions of operators with derivatives in the expansion around $(y\!-\!z)=0$, (see \eq{eq:exp3}), since they 
are suppressed by inverse powers of $m_c$. In the last equation
above we also use the  leading-power approximation (\ref{eq:masses}) and (\ref{eq:fBstat}).
Substituting in \eq{eq:corr2q} the hadronic matrix element 
(\ref{eq:hme2})  and the quark and gluon propagators in the momentum representation, we  obtain the box diagram in the form of a 
standard Feynman integral
\footnote{
Since the imaginary part
(discontinuity) of this  integral is finite, there 
is  no need for a dimensional regularization, and we put D=4.}:
\begin{eqnarray}
&&F^{(box)}_{\mu\nu}(p,q)=-4\pi\alpha_s f_{B_c}m_{B_c}
\nonumber\\
&&\times\!\!\int\!\frac{d^{4}k}{(2\pi)^{4}}
\frac{t_{\mu\nu}(p,v,k)}{\big[(m_cv-k)^2\!-\!m_c^2\big]\big[(m_cv-k-p)^2\!-\!m_c^2\big]
\big[(m_bv+k)^2\!-\!m_b^2\big]k^2}\,,
\label{eq:corr3}
\end{eqnarray}
where the dependence on $q$ is implicit via 
$v=(p+q)/M$, and the trace is reduced to 
\begin{equation}
t^{(box)}_{\mu\nu}(p,v,k)= 
2\mbox{Tr}[(-m_c\slashed{v}+\slashed{k}+m_c)\gamma_\mu(-m_c\slashed{v}+\slashed{k}+
\slashed{p}+m_c)\Gamma_\nu(m_b\slashed{v}+\slashed{k}+m_b)(2-\slashed{v})\gamma_5]\,.
\label{eq:trace}
\end{equation}
We further decompose \eq{eq:corr3} to account for the vector and axial
components of the weak current:
$$F^{(box)}_{\mu\nu}(p,v,k)=F^{V(box)}_{\mu\nu}(p,v,k)+F^{A(box)}_{\mu\nu}(p,v,k)\,,$$
 putting respectively, $\Gamma_\nu \to \gamma_\nu$ and $\Gamma_\nu\to -\gamma_\nu\gamma_5$. The trace
 (\ref{eq:trace}) splits into the two parts $t^{V(box)}_{\mu\nu}$ and $t^{A(box)}_{\mu\nu}$. 
In the case of the vector current, we find:
\begin{eqnarray}
&& 
t^{V(box)}_{\mu\nu}(p,v,k)=-8\,i\epsilon_{\mu\nu\alpha\beta}\Big(k^\alpha  p^\beta\big[m_b\!-\!m_c\!-\!2(k\cdot v)\big]
\nonumber\\
&&
+k^\alpha v^\beta\big[2m_c(k\cdot v)-k^2\big]
-v^\alpha p^\beta\big[2m_bm_c-k^2\big]\Big)\,.
\label{eq:trV}
\end{eqnarray}
A bulky expression for the trace with the axial current is presented in Appendix~\ref{app:axi}, see \eq{eq:traceA}.

The vector and axial parts of \eq{eq:corr3} with the traces 
given in  \eq{eq:trV} and \eq{eq:traceA} contain the following scalar, 
vector and tensor integrals differing by the number of 
four-momenta $k$  in the numerator:
\begin{eqnarray}
I^b_{\{1,\,\alpha,\,\alpha\beta,\,\alpha\beta(v)\}}(p^2,q^2)\equiv \!-i
\!\!\int\!\frac{ d^4k}{(2\pi)^4} \!\!\frac{ \{1,\,k_\alpha, k_\alpha k_\beta,\, k_\alpha k_\beta(k\cdot v)\}}{
k^2\big[(m_bv\!+\!k)^2\!-\!m_b^2\big]\big[(m_cv\!-\!k\!-\!p)^2\!-\!m_c^2\big]\big[(m_cv\!-\!k)^2\!-\!m_c^2\big]}
\label{eq:mastint}
\end{eqnarray}
where the index $b$ indicates the box diagram
\footnote{The global phase of the integrals is adjusted to the Cutkosky rule used in 
\eq{eq:CutkR}. }
. Furthermore, the terms  in \eq{eq:corr3} with $k^2$ in the numerator 
are reduced to the three simpler integrals  
which we denote as $J_1^b, J_\alpha^b, J_{\alpha\beta}^b $. They are 
obtained, respectively, from $I_1^b,I_\alpha^b,I_{\alpha\beta}^b$, removing $k^2$ in the denominator. We are interested only in the
imaginary part of the integrals listed above. In Appendix \ref{app:Im}, 
a detailed calculation for the scalar integral $I_1^b$ is presented, where using the
Cutkosky rule, we obtain  $\mbox{Im}I_1^b$. 
In \app{app:Tens}, decompositions of the imaginary parts of the
vector and tensor integrals are presented. The 
coefficients in these decompositions are obtained by a reduction to  scalar integrals.
We rewrite the vector-current part of the box diagram
contribution in terms of  master integrals defined in \eq{eq:mastint}:
\begin{eqnarray}
&&F^{V(box)}_{\mu\nu}(p,q)=32i\pi\alpha_s  
f_{B_c}m_{B_c}\epsilon_{\mu\nu\alpha\beta}\bigg\lbrace  I^{b\,\alpha} p^\beta(m_b-m_c) - 2   
I^{b\,\alpha\lambda}v_\lambda p^\beta +2 m_c  I^{b\,\alpha\lambda} v_\lambda v^\beta
\nonumber
\\
&&- J^{b\,\alpha} v^\beta-v^\alpha p^\beta(2 m_b m_c I_1^b-J_1^b)\bigg\rbrace \,.
\label{eq:FVbox}
\end{eqnarray}
The corresponding expression for the axial-current part is given in Appendix~\ref{app:axi}, in \eq{eq:FboxmastA}

To obtain the imaginary part of  \eq{eq:FVbox},
we use the decompositions of the imaginary parts of vector ans tensor
integrals in terms of invariant coefficients  
presented in Appendix~\ref{app:Tens}. Assuming the invariant amplitude for the vector current is defined as in \eq{eq:corr}, 
where we replace $\epsilon_{\mu\nu\alpha\beta}\,q^\alpha p^\beta=
-m_{B_c}\epsilon_{\mu\nu\alpha\beta}\,p^\alpha v^\beta$,
we finally obtain:
\begin{eqnarray}
&&\mbox{Im} F^{V(box)}(p^2,q^2)= 
-32i\pi\alpha_s f_{B_c}\big[2 m_b m_c \mbox{Im}I_1^b -(m_b-m_c)A^b -B_J^b
\nonumber\\
&&+C^b+2m_c D^b+(2m_c(p\cdot v)-p^2)F^b -2G^b\big]\,,
\label{eq:imboxFV}
\end{eqnarray}
where the expressions for $\mbox{Im}I_1^b$ and for the coefficients $A^b$, $B_J^b$\,, $C^b$, $D^b$, $F^b$, $G^b$ 
are given in \app{app:Im} and \app{app:Tens}, respectively. 
For the axial-current part, the resulting expression  for $\mbox{Im} F^{A(box)}_{\mu\nu}$  is given in  Appendix~\ref{app:axi}, in \eq{eq:resboxA}.

\subsection{The $\bar{c}c$-cut of the vertex diagram}
 The amplitude corresponding to the vertex diagram in Fig.~\ref{fig:diagNLO}(b) is derived analogously as for the box diagram, 
leading to the  complete Feynman integral:
\begin{eqnarray}
F^{(vert)}_{\mu\nu}(p,q)=\frac{4\pi\alpha_s f_{B_c}m_{B_c}}{(m_bv-q)^2-m_c^2}
\!\!\int\!\frac{d^{4}k}{(2\pi)^{4}}
\frac{t^{vert}_{\mu\nu}(p,v,k)}{\big[(k-m_cv)^2\!-\!m_c^2\big]
\big[(k+p-m_cv)^2\!-\!m_c^2\big] k^2}\,,
\label{eq:Fvert}
\end{eqnarray}
with the trace  
$$t^{(vert)}_{\mu\nu}(p,v,k)= 
\mbox{Tr}\big(\gamma ^\rho (\slashed{k}-m_c\slashed{v}+m_c)\gamma_\mu(\slashed{k}+\slashed{p}-m_c\slashed{v}+
m_c)\gamma_\rho (m_b\slashed{v}-\slashed{q}+m_c)\Gamma_\nu(1+\slashed{v})\gamma_5\big)\,.
$$
We split \eq{eq:Fvert} into the vector-current and axial-current parts:
$$F^{(vert)}_{\mu\nu}(p,v)=F^{V(vert)}_{\mu\nu}(p,v)+
F^{A(vert)}_{\mu\nu}(p,v)\,,$$
with the corresponding traces:
\begin{eqnarray}
 &&t^{V(vert)}_{\mu\nu}(p,v,k)=8\,i\epsilon_{\mu\nu\alpha\beta}\!\Big[
-2k^\alpha  p^\beta m_c
+k^\alpha v^\beta p^2 
\nonumber\\
&&+v^\alpha p^\beta\big[2m_c^2 - 2(v\cdot p)m_c 
-2(k\cdot v)m_c 
+ k^2 +2(k\cdot p)\big]\Big]
-16i\epsilon_{\rho\alpha\beta\nu}k^\rho v^\alpha p^\beta  k_\mu
\nonumber\\
&&
+16i\epsilon_{\rho\alpha\beta\mu}k^\rho v^\alpha p^\beta  v_\nu m_c
+16i\epsilon_{\rho\alpha\beta\nu}k^\rho v^\alpha p^\beta  v_\mu m_c
-16i\epsilon_{\rho\alpha\beta\nu}k^\rho v^\alpha p^\beta  p_\mu \,,
\label{eq:trvertV}
\end{eqnarray}
and $t^{A(vert)}_{\mu\nu}(p,v,k)$ presented in the Appendix~\ref{app:axi}, in \eq{eq:trvertA}.

Note  that the $c$-quark propagator multiplying the integral in \eq{eq:Fvert} 
 yields a simple pole located at the same $p^2$ value (\ref{eq:pole}) as in the LO diagram. 
 Since we are only interested in the contributions to the imaginary part of $F_{\mu\nu}^{(vert)}$
 within the duality interval (\ref{eq:dualint}) and at $q^2\lesssim 0$, this pole located 
 far above that interval yields a real-valued coefficient. 
Hence, to compute the imaginary part of \eq{eq:Fvert}, we only have to apply Cutkosky rules 
to the two $c$-quark propagators under the integral.

The full  calculation of the integral in \eq{eq:Fvert} requires 
five  integrals,
$I^v_1$, $I_\alpha^v$, I$_{\alpha\beta}^v$, $J_1^v$, and $J^v_\alpha$, 
where we use the same nomenclature as for the box diagram with the index $v$ distinguishing the
vertex diagram.
The vector-current part in terms of this notation reads:
\begin{equation}
\begin{aligned}
F^{V(vert)}_{\mu\nu}(p,q)=\frac{32i\pi\alpha_s f_{B_c}m_{B_c}}{(m_bv-q)^2-m_c^2}\bigg\lbrace\epsilon_{\mu\nu\alpha\beta}\bigg[& -2 m_c I^{v\,\alpha} p^\beta+p^2  I^{v\,\alpha} v^\beta +v^\alpha p^\beta \big(2 m_c(m_c-vp)I_1^v\\
&-2 m_cv^\lambda I_\lambda^v+J_1^v+2 p^\lambda I_\lambda^v\big)\bigg]-2 \epsilon_{\rho\alpha\beta\nu}v^\alpha p^\beta I_{\rho\mu}^v \bigg\rbrace
\end{aligned}
\label{eq:FvertV}
\end{equation}
and the expression for the axial-current part is given in Appendix~\ref{app:axi}, in \eq{eq:FvertmastA}.

Substituting  the imaginary parts of the integrals from Appendix~\ref{app:Tens}, we obtain from \eq{eq:FvertV}:
\begin{eqnarray}
\mbox{Im} F^{V(vert)}(p^2,q^2)=\frac{32i\pi\alpha_sf_{B_c}}{2m_c(p\cdot v)\!-\! p^2}\Bigg[
 2m_c\big((p\ccdot v)\!-\!m_c\big)\mbox{Im}I_1^v
 \nonumber\\
 +2(m_c \!-\!(p\ccdot v)) A^v
 -p^2 B^v +2G^v\Bigg]\,,
 \label{eq:vertIm}
\end{eqnarray} 
 where $(p\ccdot v)=p_0$ as a function of $p^2,q^2$ is taken from \eq{eq:kinem}. 
 The corresponding expression for the imaginary part of the 
axial-current contribution is in Appendix~\ref{app:axi}, in \eq{eq:resvertA}. The imaginary part of the scalar integral $\mbox{Im}I_1^v$ is derived in Appendix~\ref{app:Im}  and
expressions for all other coefficients in \eq{eq:vertIm} and in \eq{eq:resvertA} are 
obtained in Appendix~\ref{app:Tens}.

Adding together the box and vertex diagram contributions, given by \eq{eq:imboxFV} and \eq{eq:vertIm}, respectively, 
we finally obtain the OPE spectral density
\begin{equation}
\mbox{Im} F^{V(OPE)}(s,q^2)=\mbox{Im} F^{V(box)}(s,q^2)+\mbox{Im} F^{V(vert)}(s,q^2)\,,
\end{equation}
entering the sum rule (\ref{eq:VBorelSR}) for the $B_c\to J/\psi$ vector form factor
in the adopted approximation.  %

For the axial-current form factors the analogous results 
for the OPE spectral densities entering 
the sum rules (\ref{eq:A1ff}),(\ref{eq:A2ff}) and (\ref{eq:A012ff})
are obtained from:
\begin{equation}
\mbox{Im}F^{A(OPE)}_{(i)}(s,q^2)=\mbox{Im} F_{(i)}^{A(box)}(s,q^2)+
\mbox{Im} F_{(i)}^{A(vert)}(s,q^2)\,,  ~~(i)=(g), (qp), (qq)\,,
\label{eq:imAope}
\end{equation}
respectively, where all separate contributions 
are presented in Appendix~\ref{app:axi}, in Eqs~(\ref{eq:ImAopeboxg}) - (\ref{eq:ImAopevertqq}).

\section{Numerical analysis}
\label{sect:Numz}
Here we obtain  numerical results for the $B_c\to J/\psi$ 
form factors. We use the Borel sum rules
in \eqs{eq:VBorelSR}{eq:A012ff}.
Their power moments such as \eq{eq:mom} serve as a comparison.

The inputs used in our numerical analysis are collected in \Table{tab:inp}.
\begin{table}
\begin{center}
\renewcommand{\arraystretch}{1.2} 
\begin{tabular}{c c}
\toprule
Parameter & Value   \\
\midrule
 $\overline{m}_b(\overline{m}_b)$ & $ 4.18^{+0.03}_{-0.02}\, \mathrm{GeV}$\cite{Workman:2022ynf} \\
  $\overline{m}_c(\overline{m}_c)$ &$ 1.27\pm 0.02\, \mathrm{GeV}$  \cite{Workman:2022ynf}\\
   $\alpha_s(m_Z)$ & $0.1179 \pm 0.0009$ \cite{Workman:2022ynf}\\
   $\mu$ & $3.0\pm 0.5$ GeV \\
   $\alpha_s(\mu)$ & $0.2530\pm 0.0187$  \\
   $m_{B_c}$ &6.27448 $\pm$ 0.00032~ GeV \cite{Workman:2022ynf}\\
   $f_{B_c}$ & $0.434\pm0.015\,\mathrm{GeV}$  \cite{Colquhoun:2015oha}\\
   $f_{J/\psi}$ & $0.416\pm 0.06\,\mathrm{GeV}$ \cite{Workman:2022ynf} \\
$m_{J/\psi}$ & $ 3.096900\pm 0.000006 \,\mathrm{GeV}$ \cite{Workman:2022ynf} \\
${\cal M}^2$ & 3.0 $\pm$ 2.0  GeV$^2$ \\
\toprule
\end{tabular}
\caption{Input parameters used in the sum rules (\ref{eq:VBorelSR})-(\ref{eq:A012ff}).}
\label{tab:inp}
\end{center}
\end{table}
For the $b$ and $c$ quark masses, there is a certain freedom in choosing their renormalization scheme. In  the correlation function
(\ref{eq:corr}), both heavy quarks participate in two different ways: 
firstly, they form the initial $B_c$ meson in a static approximation, and, secondly, they perturbatively  propagate between  the vertices and annihilate into external currents. To account for both long-distance and short-distance dynamics of $b$ and $c$ quarks, we choose the pole scheme for their masses. 
This has certain advantages with respect to the 
$\overline{\mathrm{MS}}$ scheme which is a standard choice
for conventional QCD sum rules. Indeed, the use of the 
$\overline{\mathrm{MS}}$ masses $\overline{m}_b$ and $\overline{m}_c$ will 
bring the problem of fixing a normalization scale. Moreover,
the difference between the mass of $B_c$ and the sum 
$\overline{m}_b +\overline{m}_c$ becomes rather large to be treated as a soft scale.
We then employ the relation between the pole mass and $\overline{\mathrm{MS}}$ mass at 
$\mathcal{O}(\alpha_s)$ accuracy,
\begin{equation}
m_Q^{pole}=\overline{m}_Q(\overline{m}_Q)\left(1+\frac{4\alpha_s(\overline{m}_Q)}{3\pi}\right)\, ,~~~ Q=b,c .
\label{eq:polemass}
\end{equation}
Using for the $\overline{\mathrm{MS}}$ masses their world averages  
from \cite{Workman:2022ynf} yields:
\begin{equation}
m_b^{pole}=4.58 \pm 0.03 ~\mbox{GeV}, ~~~~
m_c^{pole}= 1.48 \pm 0.02~\mbox{GeV} \,,
\label{eq:mpole}     
\end{equation}
where we just take into account the parametric errors from  the variation of the $\overline{\mathrm{MS}}$ masses and $\alpha_s$. 
Since we work in the leading power approximation, the $\mathcal{O}(\Lambda_{QCD}/m_Q)$ 
uncertainty of the pole mass related to  long-distance effects is neglected.

To choose an optimal renormalization scale for $\alpha_s$, ideally, one should
calculate the $\mathcal{O}(\alpha_s^2)$ corrections, which are far beyond our scope and 
represent a technically challenging task for the future. 
We find useful information concerning the scale choice in Ref.~\cite{Bell:2005gw}, where gluon radiative corrections
to the one-gluon exchange were calculated in the framework of a non-relativistic
bound-state picture of the $B_c\to$ charmonium transition at large recoil. 
There it was shown that  large logarithms are absent if the scale is in the ballpark 
of $\mu\sim \sqrt{m_b m_c}$, which for the   
masses in \eq{eq:mpole}  corresponds to $\mu=2.6\pm 0.03$ GeV.  The broader interval of $\mu$ that we adopt is consistent with this prescription.

The main advantage of our sum rules is that the OPE contains 
a single non-perturbative parameter --  the $B_c$-meson decay constant. The latter can be directly measured in the leptonic decay $B_c\to \mu\nu$, which, however, has not been detected yet. We use the $f_{B_c}$ value obtained  from lattice QCD in \cite{Colquhoun:2015oha}. 
In the past, two-point QCD sum rules were also used to calculate $f_{B_c}$
(see, e.g., \cite{Colangelo:1992cx} - \cite{Kiselev:1999sc}). 
The predicted values are, within uncertainties, in agreement with the lattice QCD results. The remaining hadronic input parameters include the accurately measured masses of the $B_c$ and the $J/\psi$, 
and the decay constant of the $J/\psi$. The latter is calculated directly
from the measured leptonic width of the $J/\psi$.

Finally, we comment on the choice of the Borel mass 
interval in \Table{tab:inp}. 
In the conventional QCD sum rules based on the vacuum correlation functions and condensate expansion, two conditions have to be satisfied within this interval: \textit{i}) small power corrections, and \textit{ii})  moderate  contributions of excited and continuum states estimated using duality. An indication  that both conditions are fulfilled is the stability of the sum rule prediction with respect to the variation of $\mathcal{M}^2$.
In our setup, the conditions \textit{i}) and \textit{ii}) cannot be readily imposed, since the power corrections are neglected and the OPE spectral density 
is calculated only in the interval near $4m_c^2$. Instead, we use 
information from the well-known 2-point QCD sum rules for vector charmonium, 
which use the same interpolating current $\bar{c}\gamma_\mu c$  as the one in our correlation function.
Traditionally, starting from the original work  \cite{Shifman:1978bx}, 
the charmonium  sum rules were analysed using power moments. The latter were obtained, similarly to 
 \eq{eq:mom}, by a multiple differentiation over the external momentum square $q^2$
 and at a certain spacelike point $q^2=-P^2$. There were, however, also several analyses of the same sum rules using the Borel version, see e.g.
Refs.~\cite{Bertlmann:1981he,Beilin:1985da}, using  
the interval $1.0<\mathcal{M}^2<5.0$ GeV$^2$, which we also adopt. 
To validate this choice, we recalculated the power moments of two-point sum rules at a typical value $P^2=4m_c^2$ and at
 $n=2,3,4$. We took into account the $\mathcal{O}(\alpha_s)$ terms in the perturbative part and
the gluon condensate contribution, using the analytical expressions from
\cite{Shifman:1978bx}.  We found that both the mass and decay constant of the $J/\psi$ are 
reproduced within $\leq 5\%$ accuracy. The  moments with $n>4$ are not reliable because the 
power correction due to the gluon condensate becomes too large, whereas $n=1$ yields too large contribution 
of charmonia above the $J/\psi$ (c.f. the above criteria \textit{i}) and \textit{ii}), respectively). 
According to the approximate connection between the Borel parameter and moments, 
the optimal  set of power moments corresponds to the 
interval  2.2 GeV$^2$ $< {\cal M}^2=P^2/n< 4.4$ GeV$^2$, which is within our adopted interval in Table~\ref{tab:inp}.

The remaining element of the numerical analysis is the value of the effective threshold in the Borel sum rules. We determine it for the vector-current case, \eq{eq:VBorelSR} and use it for the other sum rules. To this end, we apply a standard tool, namely taking the derivative of the Borel sum rule over $-1/{\cal M}^2$ and dividing the result by the initial sum rule. The ratio 
 \begin{eqnarray}
\frac{\int\limits_{4m_c^2}^{s_0}\! ds \,s e^{-s/{\cal M}^2}\mbox{Im}F^{V(OPE)}(s,q^2)}{
 \int\limits_{4m_c^2}^{s_0}\! ds \,e^{-s/{\cal M}^2}\mbox{Im}F^{V(OPE)}(s,q^2)
 }=m_{J/\psi}^2\,
\label{eq:borelratio}
\end{eqnarray}
is independent of the form factor and allows us to determine $s_0$ by fitting l.h.s. to the $J/\psi$ mass.
We find that in the whole region 
$1.0~\mbox{GeV}^2 < \mathcal{M}^2< 5.0 ~\mbox{GeV}^2$  and at 
$-20.0~\mbox{GeV}^2 <q^2<0 $ 
the value $s_0=10.2$ GeV$^2$ being substituted as a threshold, yields the l.h.s of
(\ref{eq:borelratio}) equal  the squared mass of the $J/\psi$ within $\leq 2\% $.
Note that this value of $s_0$ is substantially lower than a  typical duality threshold  for the two-point sum rules which is 
close to the mass of $\psi(2S)$. This is not surprising, taking into account that here we have a completely 
different behaviour of the OPE spectral density above the
threshold $4m_c^2$.

With the chosen input, we obtain from the sum rules in
\eqs{eq:VBorelSR}{eq:A012ff} numerical values for all $B_c\to J/\psi$ form factors in the range of momentum 
transfer $-20 ~\mbox{GeV}^2< q^2<0 $. Changing the Borel parameter   within the interval in \Table{tab:inp}, we find that at $q^2=0$ the resulting variations  of all form factors at  
$\mathcal{M}^2=5.0$ GeV$^2$ ($\mathcal{M}^2=1.0$ GeV$^2$)  do not exceed 
$-2\%$ ($+6\%$ ) of their values at $\mathcal{M}^2=3.0$ GeV$^2$. Moreover, at negative $q^2<-10$ GeV$^2$ 
the form factors are practically constant with respect to the Borel mass variation. 
Thus, we reveal reasonable stability of the sum rule results, which is an important indication that the neglected power corrections are not large. 
\begin{table}
\begin{center}
\begin{tabular}{c c c c c}
\hline
Form factor & Method & $q^2 = -20\,\mathrm{GeV}^2$ &  $q^2 = -10\,\mathrm{GeV}^2$ & $q^2 = 0$  \\
\hline
$V$& (1) &0.045&0.116&0.740\\
   & (2) &0.046&0.117&0.753\\
\hline
$A_1$ & (1) &0.043& 0.093&0.472\\
& (2) & 0.043&0.094& 0.480\\
\hline
$A_2$ & (1) &0.034&0.084&0.487\\
& (2)& 0.034 & 0.084& 0.495\\
\hline
$A_0$ & (1) &0.030 &0.073&0.464\\
& (2) & 0.030 &0.074& 0.473\\
\hline
\end{tabular}
\caption{ Numerical results for the $B_c\to J/\psi$ form factors obtained from: (1) the 
Borel sum rules at ${\cal M}^2=3.0~\mbox{GeV}^2$ and (2) the power moment
sum rules with $n=3$ at $P^2=4m_c^2$. All other input parameters are taken at central values.}
\label{tab:Bor_pow}
\end{center}
\end{table}
Another test of our method is a comparison of the Borel sum rules 
with the power moments. In \Table{tab:Bor_pow} we present the form factors,
obtained from these two versions of sum rules,  setting the input parameters at their central values.
Taking $P^2/n$  maximally close to ${\cal M}^2$, we observe a good agreement, especially at 
$q^2<0$.

Our main results on the form factors are presented in \Table{tab:FF_pred}.
To obtain the  uncertainties of the form factors, we varied all the parameters in \Table{tab:inp}, apart from $\mathcal{M}^2$ to a multivariate normal distribution,  that is, uniformly in the interval $1.0\,\mathrm{GeV}^2<\mathcal{M}^2<5.0\,\mathrm{GeV}^2$. 
As a result of this procedure, we find that the predictions for the form factors do not follow a normal distribution, hence the asymmetric uncertainties in \Table{tab:FF_pred}.
The largest uncertainty, in the ballpark of $\pm 30\%$, is caused by the variation of the
$c$-quark mass, whereas the uncertainty related to 
the $b$-quark mass turns out to be at a percent level.
This  specific property of our sum rules 
is caused by the fact that the lower threshold
of the integrals depends only on $m_c$ . The second and third
in size effects influencing  uncertainties are,
respectively, the errors of $\alpha_s$ and $f_{B_c}$.

In \Table{tab:FF_pred}, we also compare our predictions at $q^2=0$ to the lattice QCD results 
\cite{Harrison:2020gvo}, finding a reasonable  agreement. Concerning the $q^2$ dependence, we find that
 the form factors based on  the new sum rules are  highly correlated, 
 preventing from  a reliable  extrapolation to the  semileptonic region $0<q^2<(m_{B_c}-m_{J/\psi})^2$.

\begin{figure}[h]
\hspace{-10mm}
    \begin{tabular}{cc} 
       \hspace{5mm} \includegraphics[width=0.45\textwidth]{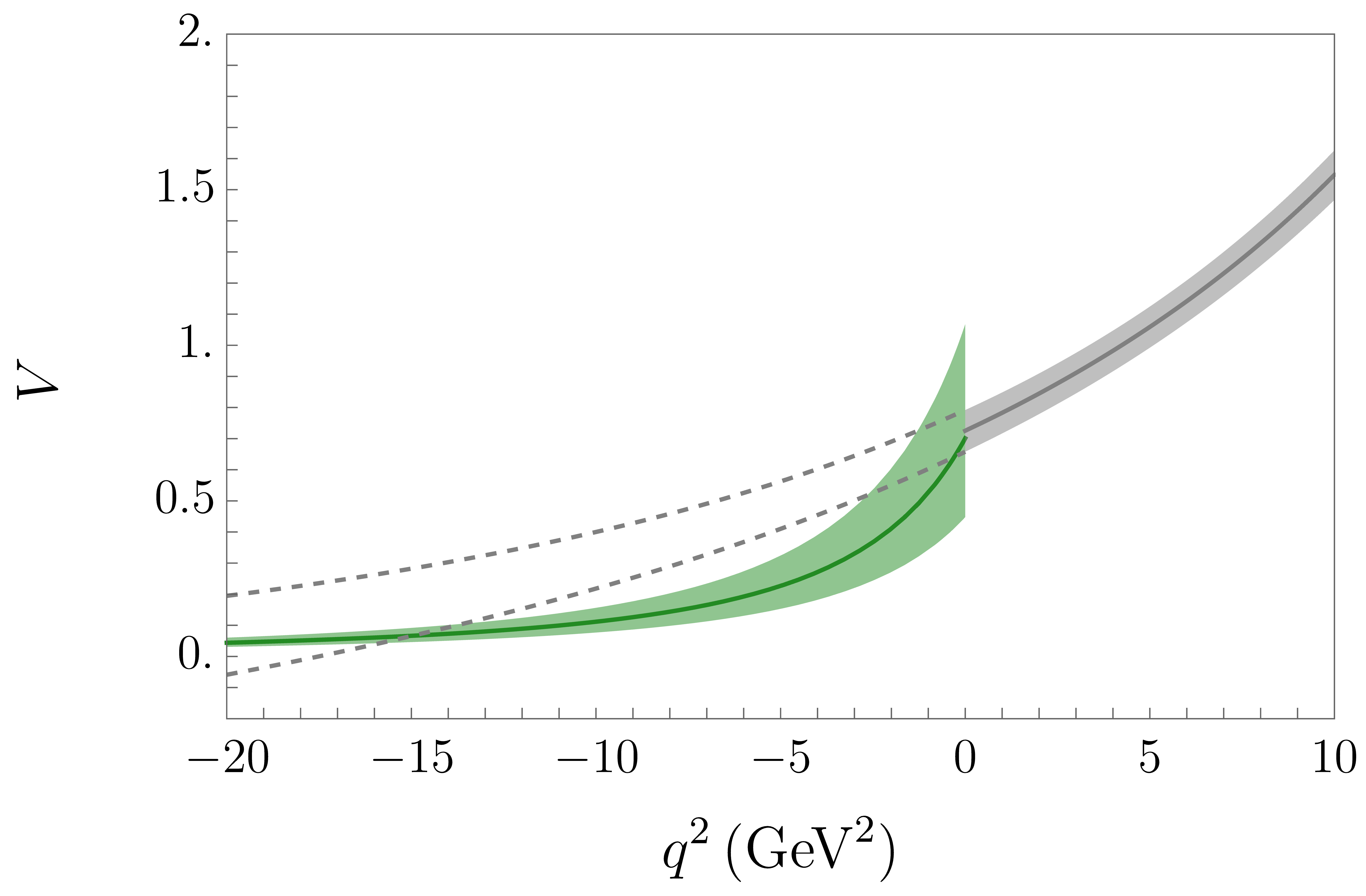} &  \hspace{5mm}
        \includegraphics[width=.45\textwidth]{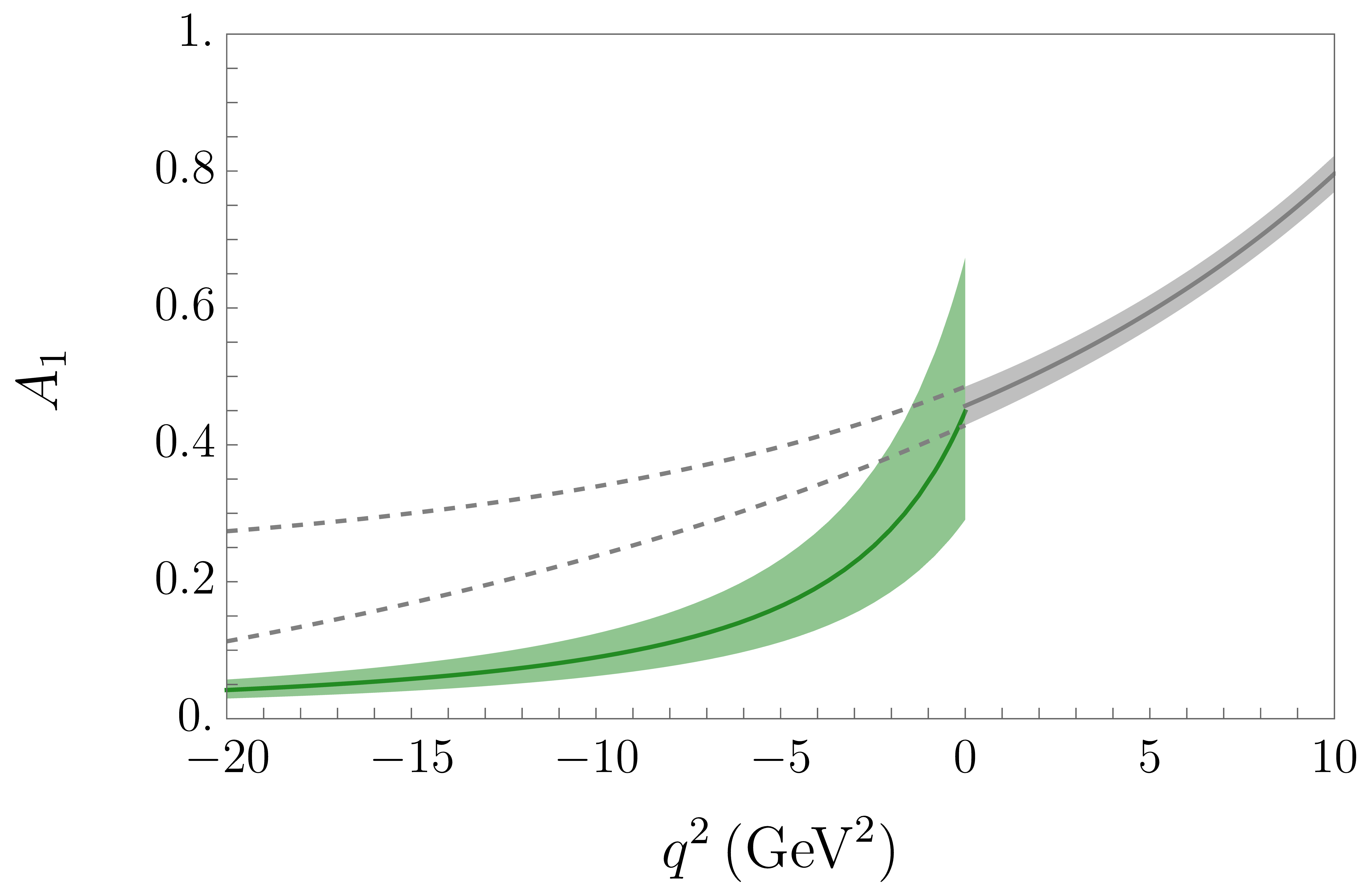}
        \\
        \hspace{5mm}\includegraphics[width=.45\textwidth]{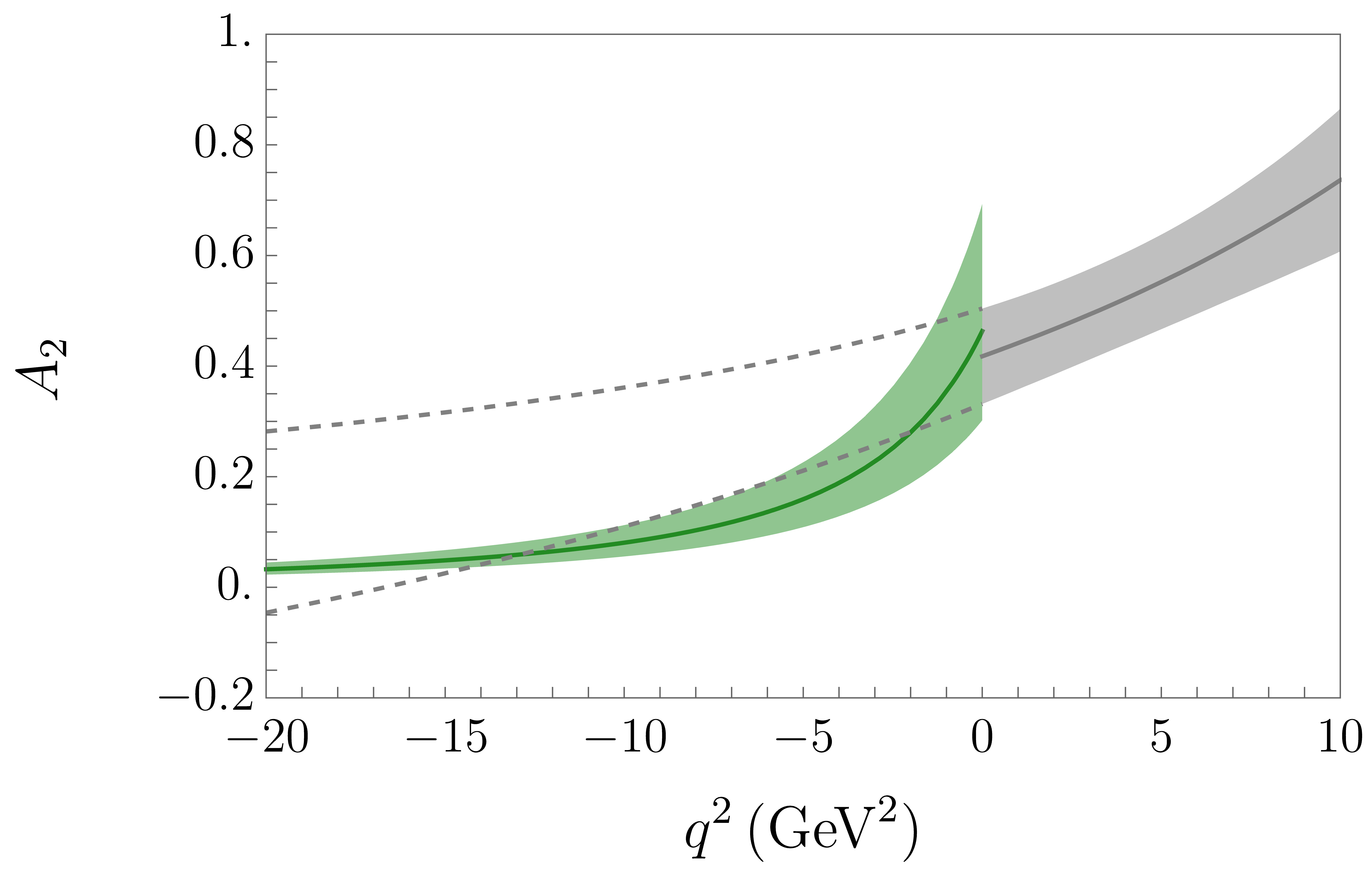} &\hspace{5mm}
        \includegraphics[width=.45\textwidth]{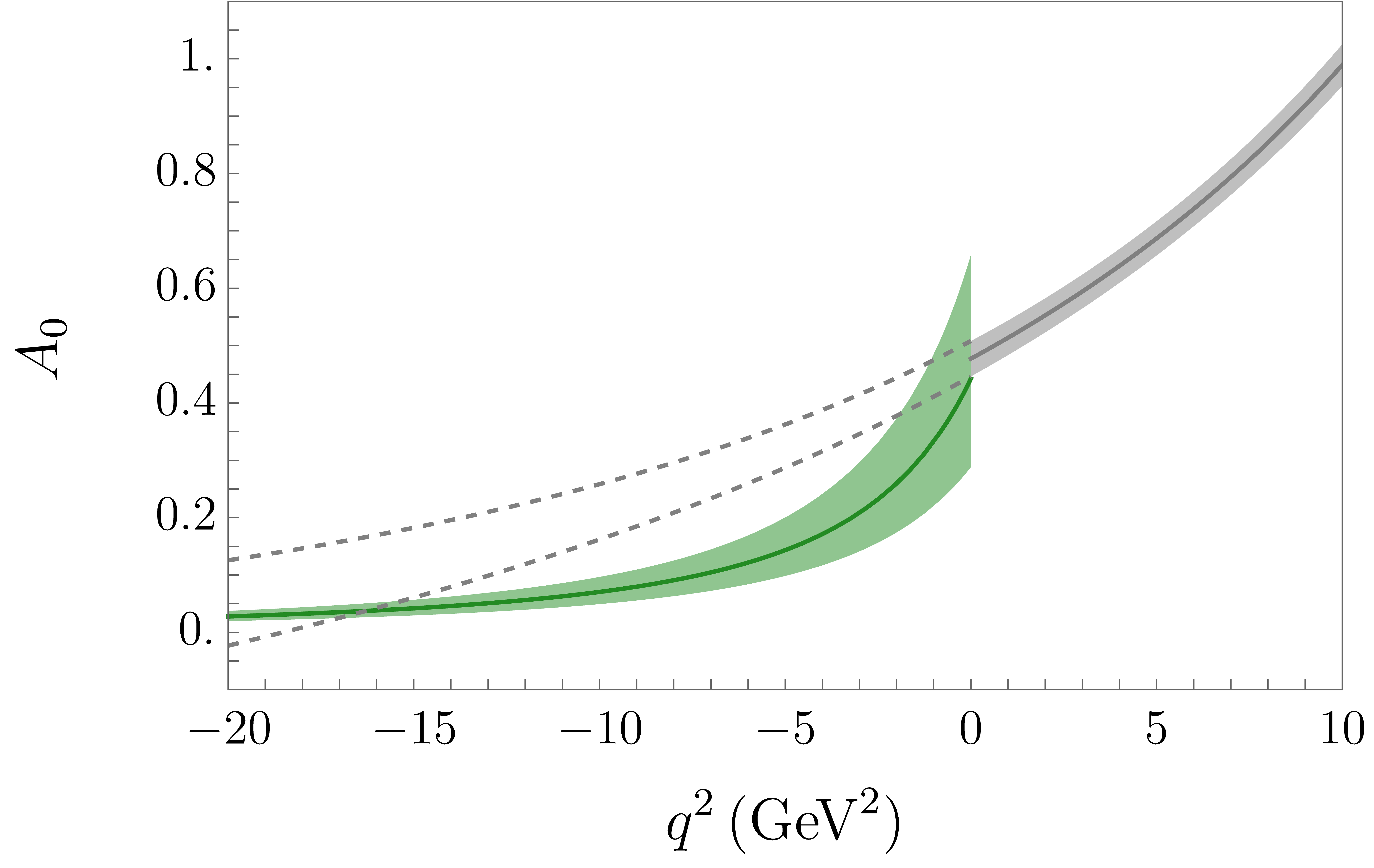}
    \end{tabular}
    \caption{Results for the vector and axial-vector $B_c\to J/\psi$ form factors. The green band represents the 68\% C.L. region for the sum rule results at $q^2<0$. The gray bands are the 68\% C.L. region for the lattice QCD predictions 
\cite{Harrison:2020gvo}    
    at $q^2>0$. The solid green and gray lines in the center show the median of the sum rules and lattice QCD results distribution, respectively.
    }
    \label{fig:FF_plots}
\end{figure}

For comparison, in \fig{fig:FF_plots}, we plot the $q^2$ dependence of the $B\to J/\psi$ form factors
obtained from the sum rules and in the lattice QCD  \cite{Harrison:2020gvo}.
 The two approaches are valid in the two complementary regions of $q^2$.
In addition, we used the $z$-expansion obtained from the lattice QCD analysis in \cite{Harrison:2020gvo}  and analytically 
continued it towards negative $q^2$. Within uncertainties, only a marginal 
agreement between the slopes is observed.
\begin{table}
\begin{center}
\renewcommand{\arraystretch}{1.2} 
\begin{tabular}{c c c c c}
\toprule
Form factor & $q^2 = -20\,\mathrm{GeV}^2$ &  $q^2 = -10\,\mathrm{GeV}^2$ & $q^2 = 0$ & HPQCD at $q^2=0$   \cite{Harrison:2020gvo} \\
\midrule
$V$ & $0.044^{+0.016}_{-0.013}$ & $0.112^{+0.043}_{-0.035}$  &  $0.705^{+ 0.364}_{-0.253}$ & $0.725\pm 0.055$ \\
$A_1$ & $0.042^{+0.015}_{-0.012}$  & $0.090^{+0.034}_{-0.027}$  & $0.451^{+0.222}_{-0.158}$ & $0.457\pm0.027$ \\
$A_0$ & $0.028^{+0.010}_{-0.008}$ & $0.071^{+0.026}_{-0.021}$ & $0.443^{+0.219}_{-0.156}$ & $0.4770\pm 0.026$ \\
$A_{2}$ & $0.033^{+0.012}_{-0.010}$  & $0.081^{
+0.031}_{-0.025}$ & $0.466^{+0.228}_{-0.162}$ & $0.418\pm 0.086$ \\
$A_{12}$ & $0.009^{+0.003}_{-0.002}$  & $0.017^{+0.006}_{-0.005}$ & $0.085^{+ 0.042}_{-0.030}$ & $0.091\pm0.008$ \\
\toprule
\end{tabular}
\caption{Numerical results for the $B_c\to J/\psi$ form factors. 
The central values (asymmetric uncertainties) correspond to the medians of the distributions (the $68\%$ confidence intervals).}
\label{tab:FF_pred}
\end{center}
\end{table}

\section{Conclusion}
\label{sect:disc}
In this paper, we derive new QCD sum rules for the 
$B_c\to J/\psi$ form factors. We use a $B_c$-to-vacuum correlation
function of the weak $b\to c$ and interpolating $\bar{c}\gamma_\mu c$
currents, retaining finite $b$ and $c$ quark masses. 
The $B_c\to J/\psi$ form  factors  are then related to the correlation function 
via the hadronic dispersion relation in the momentum squared of the interpolating current.  

We find and use two generally unexpected properties of the underlying 
correlation function, valid in the heavy-quark limit, $m_b,m_c\gg \Lambda_{QCD}$.
First, due to the presence of a heavy spectator 
$\bar{c}$-quark in $B_c$, a local 
OPE is valid in the $q^2\lesssim 0$ region of a 
large recoil to the final state.
The non-perturbative input, at the leading power in $1/m_{b,c}$ 
is reduced to a single parameter -- the decay constant of $B_c$ meson. 
Second, in the same region $q^2\lesssim 0$, the OPE spectral density in the duality interval for $J/\psi$ starts at $\mathcal{O}(\alpha_s)$. 
This is in accordance with an expected dominance of the hard-gluon exchange
in the $B_c\to J/\psi$ transition at large recoil. 

Derivation of the sum rules and calculation of the OPE diagrams 
performed in this paper is just a first step in the exploration
of the new sum rules. We foresee other important applications with various
final $\bar{c}c$ states in $B_c$ decays. 

The accuracy 
of the new sum rules can also be improved in the future. The unaccounted 
next-to-leading power corrections
proportional to the inverse $b$ and $c$ quark masses should
be identified and estimated. Their systematic analysis will demand technically 
more involved but straightforward computational procedures, leading to a couple of new input parameters. Improving also the perturbative part of the sum rules, one has to calculate the two-gluon exchanges at  $O(\alpha_s^2)$. Here the Coulomb gluons have to be factorized from the hard gluons. At this level of accuracy, the NRQCD description of the initial $B_c$ state can be used and combined with the new sum rules.

Altogether, we believe that our approach is more adequate for describing the $B_c\to$ charmonium  transitions
than the traditional framework of three-point QCD sum rules used in the past, mainly because the latter sum rules still miss important but technically inaccessible hard-gluon effects. 

Finally, we perform a numerical analysis of the leading order term.  We find that near $q^2=0$, the form factors calculated with  our method 
agree with the most advanced lattice QCD predictions.
Still, strong correlations between the sum rule numerical results at 
different $q^2<0$ points do not allow us to obtain a reliable 
extrapolation to the semileptonic region, with the help of e.g., a $z$ expansion.

Concluding, the new QCD sum rules have the potential to provide
the $B_c\to$ charmonium form factors at large recoil and can  
complement the lattice QCD results in the flavour-oriented applications of these form factors.  

\section*{Acknowledgments}
ThM thanks Matthias Steinhauser for a useful discussion on technical aspects. The research of AK and ThM was supported by the Deutsche Forschungsgemeinschaft (DFG, German Research Foundation) under grant 396021762 - TRR 257 “Particle Physics Phenomenology after the Higgs Discovery''.

\appendix

\section{Imaginary parts of the scalar integrals}
\label{app:Im}

Here we compute  the imaginary part of the scalar master integrals corresponding to the $\bar{c}c$ cut in the diagrams in \figs{fig:diagNLOa}{fig:diagNLOb}.
\subsection{Box diagram}
We start from the expression in \eq{eq:mastint} for $I_1^b$ and apply 
 Cutkosky rules to both $c$-quark propagators
\footnote{The formula for this rule  for a generic Feynman diagram can be found e.g.  in \cite{Itzykson:1980rh}.} 
: 
\begin{eqnarray}
2\mbox{Im}\,I_1^b &= & \int \frac{d^4k}{(2 \pi)^4} \frac{1}{k^2[(m_b v + k)^2-m_b^2]} 
\nonumber\\
&\times&(2\pi) \delta_+ ((m_c v - k)^2-m_c^2 ) \, (2\pi) \delta_- ((m_c v - k-p)^2-m_c^2 )\,,
\label{eq:CutkR}
\end{eqnarray} 

where for a generic on-shell momentum $P$, $P^2=m^2$, we denote  
$$
\delta_\pm (P^2-m^2) = \theta(\pm P_0)  \delta (P^2-m^2)\,.
$$
Note that for the kinematic branch we are interested in, we have to pick the antiparticle part of the 
cut charm propagator, hence the $\delta_-$ -function. 

The on-shell $\delta$-functions imply
\begin{eqnarray}
&& (m_c v - k)^2-m_c^2 = k^2 - 2 m_c (k\cdot v) = 0 \,, \label{eq:onshell1} \\
&& (m_c v - k-p)^2-m_c^2 = p^2  - 2 m_c (p\cdot v) + 2(k \cdot p)=0 \,, \label{eq:onshell2}
\end{eqnarray} 
and we can write:
\begin{eqnarray}
&& 2\mbox{Im}I^b_1 =\int \frac{d^4k}{(2 \pi)^2} \frac{1}{2 m_c (k\cdot v)} \frac{1}{2 M (k\cdot v)} 
\delta (k^2-2 m_c (k\cdot v) ) \theta(m_c-(k\cdot v)) 
\nonumber
\\
&&\qquad \times \delta( p^2  - 2 m_c (p\cdot v) + 2 (k\cdot p)) \theta((p\cdot v) + (k\cdot v) - m_c)\,, 
\label{eq:intc}
\end{eqnarray}
where $M = m_b + m_c$.

In  the reference frame (\ref{eq:frame}) of our choice
\begin{equation}
 k\cdot v=k_0\,, \quad  p\cdot v=p_0\,,  \quad k^2 = k_0^2 - \kappa^2\,, \quad p^2= p_0^2-\vec{p}^{\,2}\,,
\end{equation} 
 where $\kappa\equiv |\vec{k}|$. Hence, (\ref{eq:onshell1}) implies, 
 together with $\theta(m_c-(k\cdot v)) = \theta(m_c-k_0)$,
\begin{equation}
k_0 = m_c - \sqrt{\kappa^2+m_c^2} \equiv m_c - E_\kappa \le 0 \,. 
\end{equation}
We rewrite the $\delta$-function corresponding to \eq{eq:onshell1} as a function of $k_0$, finding:
\begin{equation}
\delta \big(k^2-2 m_c (k\cdot v) \big) \theta\big(m_c-(k\cdot v)\big) = \delta(k_0 - m_c + E_\kappa) \frac{1}{2 E_\kappa} \,.
\end{equation}
Performing the $k_0$ integration in (\ref{eq:intc}), we get
\begin{equation}
2\mbox{Im}I_1^b = \frac{1}{8 m_c M} \int \frac{d^3k}{(2 \pi)^2} \frac{1}{(m_c - E_\kappa)^2 E_\kappa} 
\delta  \big( p^2  - 2 m_c p_0 + 2 (k\cdot p)\big)
\theta\big(p_0-E_\kappa\big) \,.
\label{eq:3int}
\end{equation}
The remaining scalar product evaluated in the frame (\ref{eq:frame}) is:
\begin{equation}
k\cdot p = p_0 (m_c - E_\kappa) - \kappa |\vec{p}\,| \cos \Theta \,, 
\end{equation} 
where the angle $\Theta$ is between $\vec{p}$ and $\vec{k}$. 
The 3-momentum integration can be written as 
\begin{equation} 
\frac{d^3k}{(2 \pi)^2} = \frac{1}{2 \pi} \kappa^2 \, d \kappa \, d(\cos \Theta) 
= \frac{1}{2 \pi} \kappa^2 \, d \kappa \, du \, \theta(1-u) \theta(1+u) \end{equation}
with $u=\cos \Theta$. The remaining $\delta$-function  in \eq{eq:3int} fixes the value of the variable $u$:
\begin{eqnarray} 
p^2  - 2 m_c p_0 + 2 (k\cdot p) = p_0^2 - |\vec{p}\,|^2 - 2 m_c p_0 + 2 p_0 (m_c - E_\kappa) - 2 \kappa |\vec{p}\,| u  = 0\,,
\end{eqnarray}
so that
\begin{equation}
u_*   = \frac{1}{2 \kappa |\vec{p}\,|} 
(p_0^2 - |\vec{p}\,|^2 - 2 p_0 E_\kappa) \,,
\end{equation}
allowing us to transform the integral (\ref{eq:3int}) into
\begin{eqnarray}\label{kint}
&&2\mbox{Im}I_1^b = \frac{1}{32 \pi|\vec{p}\,| m_c M} \int \frac{\kappa \, d\kappa}{(m_c - E_\kappa)^2 E_\kappa}  
\theta(p_0-E_\kappa) \theta(\kappa) \theta(1-u_*)\theta(1+u_*) 
\nonumber \\ 
&&= \frac{1}{32 \pi |\vec{p}\,| m_c M} \int \frac{dE_\kappa}{(m_c - E)^2}
\theta(p_0-E_\kappa) \theta(E_\kappa-m_c) \theta(1-u_*)\theta(1+u_*)\,,
\end{eqnarray}
where the limits are fixed by the $\theta$ functions: 
\begin{eqnarray} 
&& 1-u_* >0: \qquad 2 |\vec{p}\,|\sqrt{E_\kappa^2-m_c^2}  - p_0^2 + |\vec{p}\,|^2 + 2 p_0 E_\kappa  > 0 \,, 
\label{eq:bound1}
\\ 
&& 1+u_* >0: \qquad 2  |\vec{p}\,|\sqrt{E_\kappa^2-m_c^2}  + p_0^2 - |\vec{p}\,|^2  - 2 p_0 E_\kappa > 0 \,.
\label{eq:bound2}
\end{eqnarray}
Note that  in the chosen frame, the momentum components $p_0$ and $|\vec{p}\,|$
are related to the variables $p^2$ and $q^2$ via \eq{eq:kinem}.

Both limits in \eqs{eq:bound1}{eq:bound2} have to be satisfied. To find them, we first look at the zeros of $(1-u_*)(1+u_*)$, that are at
\begin{eqnarray} 
E_+ = \frac{1}{2} \left[ p_0 + |\vec{p}\,|\sqrt{1- \frac{4m_c^2}{p^2}} \right] \,,~~
E_- = \frac{1}{2} \left[ p_0 - |\vec{p}\,| \sqrt{1- \frac{4m_c^2}{p^2}}\,, \right] \,,  
\label{eq:boundsE} 
\end{eqnarray}
where, obviously, $E_+ \ge E_-$. Using this, we can rewrite 
\eqs{eq:bound1}{eq:bound2} as a single inequality:
\begin{equation}
- 4 p^2 (E_\kappa-E_-)(E_\kappa-E_+) \ge 0 \,,
\end{equation}
implying for $p^2 >0$ 
\begin{equation}
(E_\kappa-E_-)(E_\kappa-E_+) \le 0 
\end{equation}
which holds for $E_- \le E_\kappa \le E_+$, defining the range of integration in \eq{kint}. 
Performing now the $E_\kappa$ integration yields,
finally, the imaginary part of the
scalar master integral (\ref{eq:mastint}) as a function of $p^2$
and $q^2$: 
\begin{equation}
\mbox{Im}I_1^b(p^2,q^2) = \frac{1}{16 \pi m_c} \sqrt{1-\frac{4 m_c^2}{p^2}} \left[\frac{p^2 M}{\big(M p^2 - m_c(M^2+p^2-q^2)\big)^2}\right]\,,
\end{equation}
where the phase space is restricted by $p^2 \ge 4 m_c^2$. Note that
the denominator has a zero at 
\begin{equation}
p^2 = \frac{m_c}{M-m_c} (M^2-q^2) 
\end{equation} 
which is at the same position as the pole 
(\ref{eq:pole}) in the LO diagram.
However, this is not relevant, since we  will stay far away from this by choosing the duality interval around $p^2\sim 4 m_c^2$. \\

\subsection{Vertex diagram}
The  scalar master integral entering the full expression (\ref{eq:Fvert})
is  simpler than for the box diagram, because the $c$-quark propagator is outside
the $k$ -integration. The Cutkosky rule yields: 
\begin{equation}
2\mbox{Im}I_1^v = \int \frac{d^4k}{(2 \pi)^4} \frac{1}{k^2}    
(2\pi) \delta_+ ((m_c v - k)^2-m_c^2 ) \, (2\pi) \delta_- ((m_c v - k-p)^2-m_c^2 )\,,
\end{equation} 
where we use the same definitions as in   the previous subsection. 
Using the on-shell relations, the reference frame (\ref{eq:frame}) and following the 
same steps in the integration procedure, we arrive at a similar integral over the variable $E_\kappa$ 
as before,  with one power of $(m_c - E_\kappa)$ less in the denominator.
The final answer reads:
\begin{equation}
\mbox{Im}I_1^v(p^2,q^2)= \frac{M}{16\pi m_c}\ell(p^2,q^2)\,,
\label{eq:Mintv}
\end{equation}
with a shorthand notation for the  combination of the logarithmic and K\"allen functions:
\begin{equation}
\ell(p^2,q^2)=\frac{1}{\sqrt{\lambda(M^2,p^2,q^2)}}
\log\bigg(\frac{M^2-4 M m_c+p^2-q^2-\sqrt{1-\frac{4 m_c^2}{p^2}} \sqrt{\lambda(M^2,p^2,q^2)}}{M^2-4 M m_c+p^2-q^2+\sqrt{1-\frac{4 m_c^2}{p^2}} \sqrt{\lambda(M^2,p^2,q^2)}}\bigg)\,.
\label{eq:log}
\end{equation}
Note that, as expected, $\ell(4m_c^2,q^2)=0$ reflecting the phase space.

\section{Scalar and tensor integrals}
\label{app:Tens}
\subsection{The box diagram}

Here we work out the integrals in (\ref{eq:mastint}) containing powers of the four-momenta $k$ in the numerator. Their imaginary parts are decomposed in all possible independent Lorentz structures:
\begin{eqnarray}
&& \mbox{Im} I_\alpha^b(p^2,q^2) =\, v_\alpha A^b + p_\alpha B^b\,,
\label{eq:Ia}\\
&&   \mbox{Im} I_{\alpha\beta}^b(p^2,q^2) =\, v_\alpha v_\beta C^b + (p_\alpha v_\beta+p_\beta v_\alpha)D^b +p_\alpha p_\beta F^b + g_{\alpha\beta}G^b\,, 
\label{eq:Iab}\\
&& \mbox{Im} I^b_{\alpha\beta(v)}(p^2,q^2)=
 v_\alpha v_\beta \widetilde{C}^b + (p_\alpha v_\beta+p_\beta v_\alpha)\widetilde{D}^b +p_\alpha p_\beta \widetilde{F}^b + 
 g_{\alpha\beta}\widetilde{G}^b\,, 
 \label{eq:Iabv}
 \end{eqnarray}
where the invariant coefficients $A^b,...,\widetilde{G}^b$ depend on $p^2$ and $q^2$. Below, for brevity we will suppress this dependence for all these coefficients and integrals.  
Contracting the above equations with different combinations of four-momenta $p,v$ and with $g_{\mu\nu}$, and solving the resulting system of linear equations, yields for the invariant coefficients in \eq{eq:Ia}: 
\begin{align}
    A^b =&\, \frac{1}{(p\cdot v)^2-p^2}\big[-p^2\big(v^\alpha\mbox{Im} I_\alpha^b\big)
     + (p\cdot v) \big(p^\alpha\mbox{Im} I_\alpha^b\big)\big]\,,
     \label{eq:Adec}
     \\
    B^b = &\, \frac{1}{(p\cdot v)^2-p^2}\big[(p\cdot v)\big(v^\alpha\mbox{Im} I_\alpha^b\big) -  
    \big(p^\alpha\mbox{Im} I_\alpha^b\big)\big]\,.
    \label{eq:Bdec}
    \end{align}
The terms in \eqs{eq:Adec}{eq:Bdec} representing contractions of four-vectors with the imaginary part of the vector integral $ I_\alpha^b$  are calculated  applying the procedure presented in \app{app:Im}. We obtain:
\begin{eqnarray}
v^\alpha \mbox{Im} I_\alpha^b(p^2,q^2)= \frac{1}{32\pi m_c }\ell(p^2,q^2)\,, 
\label{eq:vIcontr}
\end{eqnarray}
where the compact notation (\ref{eq:log}) is used, and
\begin{eqnarray}
p^\alpha\mbox{Im} I_\alpha^b=\, \frac{p^2}{32\pi m_c}\sqrt{1-\frac{4m_c^2}{p^2}} \frac{1}{[m_c(M^2+p^2-q^2)-M p^2]} \,. %
\label{eq:pIcontr}    
 \end{eqnarray}

The coefficients in the decomposition in \eq{eq:Iab} are:
\begin{align}
C^b=&\, \frac{p^2\left(
   (p\!\cdot\! v)^2\!-p^2\right) g^{\alpha \beta }+p^{\alpha } p^{\beta } (p^2+2 (p\!\cdot \!v )^2)+3 p^4 v^{\alpha } v^{\beta }-6 p^2  (p \! \cdot\!v) p^{\alpha
   }v^{\beta }}{2(p^2 -(p\ccdot v)^2)^2} \mbox{Im} I_{\alpha\beta}^b\,, 
   \label{eq:Cb}\\
D^b=&\, \frac{(p\ccdot v)(p^2 -(p\ccdot v)^2) g^{\alpha \beta }-3 (p\ccdot v) p^{\alpha } p^{\beta }-3 p^2 (p\ccdot v) v^{\alpha } v^{\beta
   }+2\left(
   p^2+2 (p\ccdot v)^2\right) p^\alpha v^{\beta }}{2(p^2 -(p\ccdot v)^2)^2}\mbox{Im} I_{\alpha\beta}^b\,, 
\label{eq:Db}\\ 
F^b=&\, \frac{\left((p\ccdot v)^2-p^2\right)
   g^{\alpha \beta }+3 p^{\alpha } p^{\beta }+\left(p^2+2
   (p\ccdot v)^2\right) v^{\alpha } v^{\beta }-6  (p\ccdot v) p^{\alpha }v^{\beta }}{2 (p^2-(p\ccdot v)^2)^2}\mbox{Im} I_{\alpha\beta}^b\,,
   \label{eq:Fb}\\
G^b=&\, \frac{\left(p^2-(p\ccdot v)^2\right) g^{\alpha \beta }-p^{\alpha } p^{\beta }-p^2 v^{\alpha }
   v^{\beta }+2 (p\ccdot v)  p^{\alpha } v^{\beta }}{2(p^2-(p\ccdot v)^2)}\mbox{Im} I_{\alpha\beta}^b(p^2,q^2)\,. 
 \label{eq:Gb}  
   \end{align}
 For the contractions entering these coefficients we obtain
\begin{align}
    g^{\alpha\beta}\mbox{Im} I_{\alpha\beta}^b=&\, \frac{1}{16\pi}\ell(p^2,q^2)\,,
    \label{eq:Ig}\\
    p^\alpha p^\beta\mbox{Im} I_{\alpha\beta}^b=&\, \frac{p^2}{64\pi M m_c}\sqrt{1-\frac{4 m_c^2}{p^2}}\,, 
    \\
v^\alpha v^\beta\mbox{Im} I_{\alpha\beta}^b=&\, \frac{1}{64\pi M m_c}\sqrt{1-\frac{4 m_c^2}{p^2}}\,,
    \\
    p^\alpha v^\beta\mbox{Im} I_{\alpha\beta}^b=&\, \frac{m_c(M^2+p^2-q^2)-M p^2}{64\pi M m_c}
    \ell(p^2,q^2)\,.
    \label{eq:Iabcontr}
\end{align}
Finally, for the decomposition (\ref{eq:Iabv}) we can use the same relations
for the invariant coefficients $\widetilde{C}^b$, $\widetilde{D}^b$, $\widetilde{F}^b$,
$\widetilde{G}^b$ as  (\ref{eq:Cb})- (\ref{eq:Gb}), replacing on r.h.s. 
$\mbox{Im} I_{\alpha\beta}^b$ by
$v^\rho\mbox{Im} I_{\alpha\beta\rho}^b$ . For the contractions we obtain:
\begin{align}
    g^{\alpha\beta}v^\rho\mbox{Im} I_{\alpha\beta\rho}^b=&\, v^\rho \mbox{Im}J_\rho^b = \frac{1}{32\pi M} \sqrt{1-\frac{4m_c^2}{p^2}}\,,
    \\
    v^\alpha v^\beta v^\rho\mbox{Im} I_{\alpha\beta\rho}^b=&\,
    -\frac{M^2-4 M m_c+p^2-q^2}{256 \pi m_c M^2}\sqrt{1-\frac{4m_c^2}{p^2}}\,, \\
    p^\alpha p^\beta v^\rho\mbox{Im} I_{\alpha\beta\rho}^b=&\,\frac{(M^2 m_c-M p^2+m_c(p^2-q^2))^2}{128 \pi m_c M^2} \ell (p^2,q^2)\,,
    \\
    p^\alpha v^\beta v^\rho\mbox{Im} I_{\alpha\beta\rho}^b=&\,\frac{M^2 m_c-M p^2+m_c(p^2-q^2)}{128 \pi m_c M^2} \sqrt{1-\frac{4m_c^2}{p^2}}\,. 
\end{align}
We now focus on the J-type integrals.
For the scalar integral $J_1^b$, we use the relation 
$$\mbox{Im} J_1^b = g^{\alpha\beta} \mbox{Im} I^b_{\alpha\beta}\,,$$
reducing it to the contraction (\ref{eq:Ig}).
For the vector integral, we use the decomposition:
\begin{align}
    \mbox{Im}J_\alpha^b(p^2,q^2) =&\, v^\alpha  A_J^b+ p^\alpha B_J^b\,,
\end{align}
where the expressions for the coefficients $A_J^b,\, B_J^b$ are 
obtained from (\ref{eq:Adec}) and (\ref{eq:Bdec}) replacing on r.h.s.
$\mbox{Im} I_\alpha^b$ by $\mbox{Im} J_\alpha^b$, and the contractions are
\begin{align}
     p^\alpha \mbox{Im}J_\alpha^b=&\,\frac{m_c(M^2+p^2-q^2)-M p^2}{32\pi M }\ell(p^2,q^2)\,,\\
    v^\alpha  \mbox{Im}J_\alpha^b=&\,\frac{1}{32 \pi  M}\sqrt{1-\frac{4 m_c^2}{p^2}}\,.
\end{align}

\subsection{The vertex diagram}

To calculate the imaginary part of the vector and tensor integrals,
we use decompositions similar to the ones in \eqs{eq:Ia}{eq:Iabv}:
\begin{align}
  \mbox{Im} I_\alpha^v =&\, v_\alpha A^v + p_\alpha B^v\,,
  \label{eq:Iav}\\  
\mbox{Im} I_{\alpha\beta}^v =&\, v_\alpha v_\beta C^v + (p_\alpha v_\beta+p_\beta v_\alpha)D^v +p_\alpha p_\beta F^b + g_{\alpha\beta}G^v\,.
\label{eq:Iabvert}  
\end{align}
The coefficients of the tensor structures in \eqs{eq:Iav}{eq:Iabvert} are  given by the same relations in \eq{eq:Adec}, \eq{eq:Bdec}, and  \eqs{eq:Cb}{eq:Gb}, respectively,
where the index $b$ should be replaced by the index $v$.
We calculate the corresponding contractions employing the same method as 
for the scalar master integral for the vertex diagram (see \app{app:Im}). 
The results are:
\begin{align}
    v^\alpha \mbox{Im}I_\alpha^v =& \frac{1}{32\pi m_c} \sqrt{1-\frac{4 m_c^2}{p^2}}\,, \\
p^\alpha\mbox{Im} I_\alpha^v=& \frac{m_c(p^2-q^2+M^2)-Mp^2}{32\pi m_c}\ell(p^2,q^2)\,,\\
    g^{\alpha\beta} \mbox{Im} I_{\alpha\beta}^v =& \frac{1}{16\pi }\sqrt{1-\frac{4m_c^2}{p^2}}\,, \\
    p^{\alpha}p^{\beta} \mbox{Im} I_{\alpha\beta}^v =& \frac{(m_c(M^2+p^2-q^2)-M p^2)^2}{64\pi M m_c}\ell(p^2,q^2)\,,\\
     v^{\alpha}v^{\beta} \mbox{Im} I_{\alpha\beta}^v = &\frac{4 M m_c -(M^2+p^2-q^2)}{128\pi Mm_c}\sqrt{1-\frac{4m_c^2}{p^2}}\,,   \\
    p^{\alpha}v^{\beta} \mbox{Im} I_{\alpha\beta}^v =& \frac{m_c(M^2+p^2-q^2)-M p^2}{64\pi M m_c}
    \sqrt{1-\frac{4m_c^2}{p^2}}\,.  \\
\end{align}
In addition, for the scalar integral $J_1^v$  we use the relation 
\begin{equation}
\mbox{Im} J_1^v = 2m_c v^\alpha \mbox{Im}I_\alpha^v=
2m_c\big(A^v+B^v(p\cdot v)\big)\,,
\end{equation}
which follows from the on-shell conditions. Finally,
there is also a simple relation for 
the tensor integral entering the vector part:  
\begin{equation}
\epsilon_{\rho\alpha\beta\nu}\mbox{Im}I_{\mu}^{v\,\rho}p^\alpha v^\beta = \epsilon_{\mu\nu\alpha\beta}p^\alpha v^\beta G^v\,.
\end{equation}

\section{Expressions for the axial-current part }
\label{app:axi}
\begin{itemize}
\item The trace of the box diagram:
\begin{align}
\label{eq:traceA}
t^{A(box)}_{\mu\nu}(p,v,k)&= 8\Bigg[
 g_{\mu\nu}\Big((p\ccdot v) k^2  - 2(p \ccdot v)m_b m_c - 
         2(k\ccdot v)(k\ccdot p) \nonumber \\         
&- (k\ccdot v)k^2 - 2(k\ccdot v)m_b m_c +2(k\ccdot v)^2m_c + 
(k\ccdot p) (m_b-m_c) + k^2 m_b\Big) \nonumber\\
&  +(k_\mu p_\nu+k_\nu p_\mu+ 2k_\mu k_\nu)\Big( m_c - m_b +2(k\ccdot v)  \Big)\\
&+v_\mu k_\nu \Big(2m_b m_c - 2m_c^2-2(k\ccdot v)m_c - k^2 \Big)
+v_\nu k_\mu\Big(4m_b m_c - 2(k\ccdot v)m_c  - k^2\Big) \nonumber\\
&+v_\mu v_\nu m_c\Big(2k^2 -4m_b m_c\Big)
+(v_\mu p_\nu+p_\mu v_\nu)\Big(2m_bm_c-k^2\Big)\Bigg] \nonumber\,.
\end{align}

\item{The trace of the vertex diagram:}
\begin{equation}
\begin{aligned}
 t^{A(vert)}_{\mu\nu}(p,v,k)&= 8\Bigg[
 g_{\mu\nu}\Big[2(k\ccdot v)(p\cdot v)m_c  - 
 2(k\cdot p)(p\cdot v) \\
 &-k^2(p\ccdot v) +(k\ccdot v)p^2 + 2m_c(p\ccdot v)^2 - m_c p^2 \Big]\\
  & +k_\mu k_\nu\Big[ 2(p\ccdot v)
    \Big]
  -k_\mu p_\nu\Big[ 2(k\ccdot v)+2 m_c\Big] 
  +k_\nu p_\mu\Big[ 2(p\ccdot v)\Big] \\
  & +v_\mu k_\nu\Big[ - p^2 -2(p\ccdot v)m_c \Big]\\
& +v_\nu k_\mu\Big[ -p^2 + 4m_c^2 +2(p\ccdot v)m_c  
+4(k\ccdot v)m_c -2(k\ccdot p) \Big] \\
&+v_\mu v_\nu\Big[ 2m_c p^2 - 4 m_c^3 -2m_c k^2 \Big]
+v_\nu p_\mu \Big[2m_c^2-2(p\ccdot v)m_c+k^2\Big]\\
&+v_\mu p_\nu \Big[2m_c^2-2(p\ccdot v)m_c+2(k\ccdot p)+k^2 
\Big] -2p_\mu p_\nu (k\ccdot v)
\Bigg]\,.
\label{eq:trvertA}
\end{aligned}
\end{equation}

\item Decomposition of the correlation function in terms of scalar, vector and tensor integrals for the box diagram:
\begin{equation}
\begin{aligned}
&F^{A(box)}_{\mu\nu}(p,q)=-32 \pi \alpha_s f_{B_c}m_{B_c}\bigg\lbrace g_{\mu\nu}\bigg[((p\ccdot v)+m_b)J_1^b-2 (p \ccdot v)m_b m_c I_1^b \\
&-v^\alpha(2p^\beta I_{\alpha\beta}^b+J_\alpha^b+2 m_b m_c I^b_\alpha-2 m_c v^\beta I_{\alpha\beta}^b)+(m_b-m_c)p^\alpha I_\alpha^b\bigg]\\
&+m_c v_\mu v_\nu \bigg[2 J_1^b -4 m_b m_c I_1^b\bigg]
+(v_\mu p_\nu+v_\nu p_\mu)\bigg[2m_bm_c I_1^b-J_1^b\bigg] \\
&+v_\mu \bigg[2 m_c(m_b-m_c) I_\nu^b-2 m_c v^\alpha I_{\alpha\nu}-J_\nu^b\bigg]+v_\nu\bigg[4 m_b m_c I_\mu^b-2 m_c v^\alpha I_{\alpha\mu}^b-J_\mu^b\bigg] \\
&+p_\nu \bigg[(m_c-m_b)I_\mu^b+2 v^\alpha I_{\alpha\mu}^b\bigg]+p_\mu\bigg[(m_c-m_b)I_\nu^b+2v^\alpha I_{\alpha\nu}^b\bigg]\\
&+2I_{\mu\nu}^b(m_c-m_b)+ 4 v^\alpha I_{\alpha\mu\nu}^b
\bigg\rbrace\,,
\label{eq:FboxmastA}
\end{aligned}
\end{equation}

\item Decomposition of the correlation function in terms of scalar, vector and tensor integrals for the vertex diagram:
\begin{equation}
\begin{aligned}
&F^{A(vert)}_{\mu\nu}(p,q)=\frac{32\pi\alpha_s f_{B_c}m_{B_c}}{(m_bv-q)^2-m_c^2}\bigg\lbrace g_{\mu\nu} \bigg[ v^\alpha (2 m_c(p\ccdot v)+p^2)I_\alpha^v-2 (p\ccdot v)p^\alpha I_\alpha^v\\ 
&-(p\ccdot v)J_1^v+m_c(2(p\ccdot v)^2-p^2)I_1^v \bigg]
+v_\mu v_\nu \bigg[2m_c(p^2-2 m_c^2)I_1^v-2 m_c J_1^v\bigg]
\\
&+v_\mu p_\nu \bigg[2 m_c(m_c-(p \ccdot v))I_1^v+2 p_\alpha I_\alpha^v+J_1^v\bigg] \\
&+v_\nu p_\mu \bigg[2m_c (m_c-(p \ccdot v))I_1^v+J_1^v \bigg] 
-2 p_\mu p_\nu v^\alpha I_\alpha^v
\\
&+2(p\ccdot v)p_\mu I_\nu^v-2p_\nu \bigg[v^\alpha I_{\mu\alpha}^v+m_c I_\mu^v\bigg]-(p^2+2 m_c (p \ccdot v))v_\mu I_\nu^v\\
&+ v_\nu \bigg[(2 m_c (p \ccdot v)-p^2 +4 m_c^2)I_\mu^v+2(2 m_c v^\alpha-p^\alpha) I_{\mu\alpha}^v\bigg]
+2 (p \ccdot v)I_{\mu\nu}^v \bigg\rbrace\,.
\end{aligned}
\label{eq:FvertmastA}
\end{equation}

\item Imaginary part of the correlation function for the axial-current part of the box diagram
in terms of the coefficients calculated in Appendix B:
\begin{equation}
\begin{aligned}
&\mbox{Im} F_{\mu\nu}^{A(box)}(p,q) = -16 \pi \alpha_s f_{B_c} m_{B_c}\bigg\lbrace 
2g_{\mu\nu}\bigg[-(p\ccdot v) \bigg(2m_b m_c \mbox{Im}I_1^b +B_J^b\bigg)-A_J^b
\\
&+ \bigg((m_b-m_c) (p\ccdot v) -2 m_b m_c\bigg)A^b
+ \bigg(p^2 (m_b-m_c)-2m_b m_c\, (p \ccdot v)\bigg)B^b
\\ 
&  + \bigg(m_b+2 m_c-p\ccdot v\bigg)C^b
+\bigg(2 (m_b+2m_c) (p\ccdot v) -2 p^2\bigg)D^b
\\
&
+ \bigg(m_b p^2+2m_c(p \ccdot v)^2-p^2 (p \ccdot v) \bigg)F^b
+2 \bigg(m_b+2 m_c+(p\ccdot v)\bigg)G^b+
4 \tilde{G}^b\bigg]
\\
&+4 v_\mu v_\nu\bigg[ -2m_b m_c^2 \mbox{Im}I_1^b-A_J^b+(3 m_b-m_c)m_c A^b - m_b C^b 
+ m_c p^2 F^b+2 m_c G^b+2\tilde{C}^b\bigg] 
\\
&+2v_\mu p_\nu \bigg[2 m_b m_c \mbox{Im}I_1^b-(m_b-m_c)A^b+2(m_b-m_c)m_c B^b-B_J^b+C^b
\\
&-2m_b D^b  - \bigg(p^2+2 m_c (p\ccdot v) \bigg)F^b -2 G^b+4 \tilde{D}^b\bigg] 
\\
&+2p_\mu v_\nu \bigg[2 m_b m_c \mbox{Im}I_1^b-(m_b-m_c)A^b+4 m_b m_c B^b-B_J^b+C^b
\\
&-2m_b D^b - \bigg(p^2 +2 m_c (p \ccdot v))F^b-2 G^b + 4 \tilde{D}^b\bigg] 
\\
& + 4p_\mu p_\nu \bigg[(m_c-m_b)B^b+2 D^b+\bigg(m_c-m_b+2 p.v\bigg)F^b+2\tilde{F}^b\bigg) \bigg]\,.
\label{eq:resboxA}
\end{aligned}   
\end{equation}

\item The same as above, but for the vertex diagram:
\begin{equation}
\begin{aligned}
    &\mbox{Im} F^{A(vert)}_{\mu\nu}(p,q)= \frac{-16 \pi\alpha_s f_{B_c}m_{B_c}}{2m_c(p\cdot v)- p^2}
    \bigg\{ g_{\mu\nu}\bigg[
    -2m_c\left(p^2-2 (p\ccdot v)^2\right)\mbox{Im}I_1^v \\
    &+2 \left(p^2-2 (p\ccdot v)^2\right)A^v-2 p^2 (p\ccdot v)B^v +4 (p\ccdot v) G^v \bigg] \\
   & +2v_\mu v_\nu \bigg[
   2m_c \bigg(p^2-2m_c^2\bigg)\mbox{Im}I_1^v -2 p^2 A^v -4 m_c^2 ( p\ccdot v)B^v\\
   &+4 m_c C^v+\bigg(4 m_c( p\ccdot v)-2 p^2\bigg)D^v +4m_c G^v\bigg]  \\
   &+2p_\nu v_\mu \bigg[ 2m_c \bigg(m_c-(p\ccdot v)\bigg )\mbox{Im}I_1^v
   +2 (p\ccdot v)A^v+ p^2 B^v-2C^v-2 G^v\bigg]\\
    &+2p_\mu v_\nu\bigg[
    +2m_c\bigg(m_c-(p\ccdot v)\bigg) \mbox{Im}I_1^v 
  +2 \bigg(m_c+(p\ccdot v)\bigg) A^v   
+\bigg(4m_c^2+4m_c (p\ccdot v)-p^2\bigg)B^v
\\
&+4m_c D^v-2 G^v+ 2\bigg(2m_c ( p\ccdot v)-p^2\bigg)F^v \bigg ]
\\
&+4p_\mu p_\nu \bigg[- m_c B^v- A^v- D^v\bigg]  
   \bigg\}\,.
   \label{eq:resvertA}
\end{aligned}
\end{equation}

\end{itemize}

In \eq{eq:resboxA} and \eq{eq:resvertA} the OPE result for the axial-current part is obtained in terms of the 
Lorentz decomposition containing the four-vectors $v,p$: 
\begin{equation}
\begin{aligned}
    F_{\mu\nu}^A(p,q) =\,&g_{\mu\nu}F_{(g)}^A(p^2,q^2) +v_\mu v_\nu F_{(vv)}^A(p^2,q^2)+
    v_\mu p_\nu F_{(vp)}^A(p^2,q^2) \\
    +\,&v_\nu p_\mu F_{(pv)}^A(p^2,q^2)
+p_{\mu}p_{\nu}F_{(pp)}^A(p^2,q^2)\,.
\label{eq:FAv1}
\end{aligned}
\end{equation}
In order to match to \eq{eq:FAdecomp}, we replace the momentum $v$ with $q$, using the approximation $v=(p+q)/M$:
\begin{equation}
\begin{aligned}
F_{\mu\nu}^A(p,q) =\,&g_{\mu\nu}F_{(g)}^A(p^2,q^2)
+q_\mu p_\nu\bigg( \frac{F_{(vv)}^A(p^2,q^2)}{M^2}+\frac{F_{(vp)}^A(p^2,q^2)}{M}\bigg) \\
+&q_{\mu}q_{\nu}\bigg(\frac{F_{(vv)}^A(p^2,q^2)}{M^2}\bigg)
+\dots\,,
\label{eq:FAq2}
\end{aligned}
\end{equation}
where the structures that are inessential for the sum rules are not shown.
We obtain  for the box-diagram and vertex-diagram parts
of the OPE spectral function (\ref{eq:imAope}) entering the sum rule (\ref{eq:A1ff}):
\begin{align}
\mbox{Im}F^{A(box)}_{(g)}(p^2,q^2)=&
-32 \pi \alpha_s f_{B_c} m_{B_c}
\bigg[-(p\ccdot v) \bigg(2m_b m_c \mbox{Im}I_1^b +B_J^b\bigg)-A_J^b
\nonumber\\
&+ \bigg((m_b-m_c) (p\ccdot v) -2 m_b m_c\bigg)A^b
+ \bigg(p^2 (m_b-m_c)-2m_b m_c\, (p \ccdot v)\bigg)B^b
\nonumber\\ 
&  + \bigg(m_b+2 m_c-p\ccdot v\bigg)C^b
+\bigg(2 (m_b+2m_c) (p\ccdot v) -2 p^2\bigg)D^b
 \nonumber \\
&
+ \bigg(m_b p^2+2m_c(p \ccdot v)^2-p^2 (p \ccdot v) \bigg)F^b
+2 \bigg(m_b+2 m_c+(p\ccdot v)\bigg)G^b+
4 \tilde{G}^b\bigg]\,,\label{eq:ImAopeboxg} \\
\mbox{Im}F^{A(vert)}_{(g)}(p^2,q^2)=&\frac{-32 \pi\alpha_s f_{B_c} m_{B_c}}{2m_c(p\cdot v)- p^2}\bigg[
-m_c\left(p^2-2 (p\ccdot v)^2\right)\mbox{Im}I_1^v \nonumber \\
&+\left(p^2-2 (p\ccdot v)^2\right)A^v- p^2 (p\ccdot v)B^v +2 (p\ccdot v) G^v \bigg]\,,
\label{eq:ImAopevertg}
\end{align}
For the sum rule in \eq{eq:A2ff} we have: 
\begin{align}
\mbox{Im}F^{A(box)}_{(qp)}(p^2,q^2)=&
-32 \pi \alpha_s f_{B_c}m_{B_c}\Bigg\{\frac2{M^2} \bigg[ -2m_b m_c^2 \mbox{Im}I_1^b
-A_J^b+(3 m_b-m_c)m_c A^b 
\nonumber\\
&- m_b C^b + m_c p^2 F^b+2 m_c G^b+2\tilde{C}^b\bigg]
+\frac1{M}\bigg[2 m_b m_c \mbox{Im}I_1^b-(m_b-m_c)A^b
\nonumber\\
&+2(m_b-m_c)m_c B^b-B_J^b+C^b
-2m_b D^b  - \bigg(p^2+2 m_c (p\ccdot v) \bigg)F^b\nonumber\\
&-2 G^b+4 \tilde{D}^b\bigg] \Bigg\}\,,
\label{eq:ImAopeboxqp} \\
\mbox{Im}F^{A (vert)}_{(qp)}(p^2,q^2)=&\frac{-32 \pi\alpha_s f_{B_c}m_{B_c}}{2m_c(p\cdot v)- p^2}\bigg\{\frac1{M^2} \bigg[
   2 m_c \bigg(p^2-2m_c^2\bigg)\mbox{Im}I_1^v -2 p^2 A^v 
   \nonumber\\
 &  -4 m_c^2 ( p\ccdot v)B^v+4 m_c C^v+\bigg(4 m_c( p\ccdot v)-2 p^2\bigg)D^v +4m_c G^v\bigg] 
\nonumber \\
   &+\frac{1}{M}\bigg[ 2 m_c \bigg(m_c-(p\ccdot v)\bigg ) \mbox{Im} I_1^v
   +2 (p\ccdot v)A^v+ p^2 B^v-2C^v-2 G^v\bigg] \bigg\}\,,
\label{eq:ImAopevertqp}
\end{align}
and  for the sum rule in \eq{eq:A012ff}:
\begin{align}
\mbox{Im}F^{box,A}_{(qq)}(p^2,q^2)=&-\frac{64 \pi \alpha_s f_{B_c} m_{B_c}}{M^2}
\bigg[ -2m_b m_c^2 \mbox{Im}I_1^b-A_J^b+(3 m_b-m_c)m_c A^b 
\nonumber\\
&- m_b C^b + m_c p^2 F^b+2 m_c G^b+2\tilde{C}^b\bigg]\,, 
\label{eq:ImAopeboxqq} \\
\mbox{Im}F^{vert, A}_{(qq)}(p^2,q^2)=&\frac{-32 \pi\alpha_s f_{B_c}m_{B_c}}{(2m_c\big(p\cdot v)- p^2\big)M^2} \bigg[
   2 m_c \bigg(p^2-2m_c^2\bigg) \mbox{Im}I_1^v -2 p^2 A^v 
   \nonumber\\
   &-4 m_c^2 ( p\ccdot v)B^v+4 m_c C^v+\bigg(4 m_c( p\ccdot v)-2 p^2\bigg)D^v +4m_c G^v\bigg] \,.
\label{eq:ImAopevertqq}
\end{align}

\bibliographystyle{utphys}
\bibliography{references}

\end{document}